\documentclass[twocolumn,superscriptaddress,10pt,aps,prb]{revtex4-2}
\usepackage{epsfig,amsopn}
\usepackage{graphicx}
\usepackage{epstopdf}
\usepackage{braket}
\usepackage{multirow}
\usepackage{amsmath}
\usepackage{tikz}

\usetikzlibrary{patterns}
\usepackage{amsmath,amssymb}
\usepackage{amsthm}
\usepackage{enumerate}
\usepackage{xcolor}
\usepackage{bm}
\usepackage{subfigure}
\usepackage{fontawesome5}
\usepackage{enumitem}
\usepackage[colorlinks=true,linkcolor=blue,allcolors=blue,
filecolor=black,citecolor = magenta,urlcolor=magenta]{hyperref}
\definecolor{orcidlogocol}{rgb}{0.65, 0.807, 0.223}
\newcommand{\orcid}[1]{$\,$\href{https://orcid.org/#1}{\textcolor{orcidlogocol}{\faOrcid}}}

\usepackage[defaultcolor=blue]{changes}

\definecolor{Purple}{RGB}{90, 180, 90}
\newcommand{\sat}[1]{{\color{black}  #1}}
\definecolor{orange}{rgb}{1,0.5,0}

\newcommand\beq{\begin{equation}}
	\newcommand\eeq{\end{equation}}
\newcommand\bes{\begin{subequations}}
	\newcommand\ees{\end{subequations}}
\newcommand\bea{\begin{eqnarray}}
	\newcommand\eea{\end{eqnarray}}
\newcommand\non{\nonumber}
\newcommand\noi{\noindent}

\newcommand{\RNum}[1]{\uppercase\expandafter{\romannumeral #1\relax}}

\newcommand\rk[1]{{\color{black}#1}}

\begin{document}

\title{
Anomalous topological phases in a Chern insulator connected to leads in a cylindrical geometry}
\author{Satyam Sinha}
\affiliation{International Center for Theoretical Sciences (ICTS), Bengaluru 560089, India}
\affiliation{Department of Physics, Indian Institute of Science Education and Research, Pune 411008, India}
\author{Rekha Kumari}
\affiliation{International Center for Theoretical Sciences (ICTS), Bengaluru 560089, India}
\affiliation{Advanced Institute for Materials Research (WPI-AIMR), Tohoku University, Sendai 980-8577, Japan}
\author{Junaid Majeed Bhat}
\affiliation{University of Kashmir, Kupwara Campus, Wayen, Kupwara 193222, India}
\author{Abhishek Dhar}
\affiliation{International Center for Theoretical Sciences (ICTS), Bengaluru 560089, India}
\author{Diptiman Sen}
\affiliation{Centre for High Energy Physics, Indian Institute of Science, Bengaluru 560012, India}
\author{R. Shankar}
\affiliation{The Institute of Mathematical Sciences, CIT Campus, Chennai 600113, India}

\begin{abstract}

The observed robustly quantized Hall conductance in quantum Hall systems and Chern insulators (CI) is normally  understood in terms of the bulk topology of isolated systems, not coupled to leads. It is assumed that the leads act as inert reservoirs. Within a model of a CI
coupled to leads with a cylindrical geometry, we show that this is not always true.  We identify the  Hall conductance with a boundary invariant, the winding number of the phase of the reflection coefficient.  We find anomalous topological phases where the boundary invariant is not  the same as the bulk Chern number, even in the limit of weak lead coupling. 
\end{abstract} 
\maketitle

\textit{Introduction:}
Insulators with non-trivial topology, \rk{including} quantum Hall systems, Chern insulators (CI) and time-reversal-invariant topological insulators support edge modes that carry dissipationless currents~\cite{halperin1982,ref_bernevig2013,moore2010,hasan2010, qi2011}. 
The idea of topological phases of quantum matter originated from the observation of the integer quantum Hall effect (IQHE) ~\cite{vonKlitzing1980}. The attempt to understand
why the IQHE Hall conductance plateaus are so
robustly quantized in units 
of $e^2/h$ led to the concept
of topological phases. In a seminal work, Laughlin~\cite{laughlin1981quantized} presented a very general argument, based on the principle of gauge invariance, that when the Fermi level $\mu$ is in the gap, the Hall conductance, $G_H(\mu)$
has to be quantized as $G_H(\mu)=ne^2/h$, where $n$ is an unspecified integer. This argument was made within a cylindrical geometry and the so-called charge pumping setup. 

Thouless and others \cite{thouless} (TKNN)
considered the bulk system with periodic boundary conditions and showed, using the Green-Kubo formalism, that the Hall
\emph{conductivity} was quantized as $\sigma_H(\mu)=\frac{e^2}{h}\mathcal{C}$, where 
$\mathcal{C}$ is an integer-valued topological invariant, the Chern number.

However, experiments typically measure the Hall conductance in a system with edges. The Chern number, a 
bulk quantity, defined in the torus geometry, was 
related to a topological invariant of 
the isolated system in the cylindrical/Corbino geometry 
(systems with edges) by several works \cite{halperin1982,hatsugai1993,dwivedi2016},
which showed that $\mathcal{C}$ is equal
to the number of edge channels $n$ which intersect the Fermi level. This is the
so-called bulk-boundary correspondence (BBC).  The BBC is experimentally probed 
by coupling leads to the CI and measuring the Hall 
conductance. If we assume that the leads are just
passive sources/sinks of electrons and that the physics is controlled by the spectrum of the isolated system with edges, then it can be argued that $G_H(\mu)=n=\mathcal{C}$, \rk{in units of $e^2/h$}~\cite{buttiker1988}. 

In this paper we show
that this is not always true within
a microscopic model of the leads and their coupling to the CI. We analyse a simple exactly solvable
model; we show that, while $G_H$ is always quantized, there are phases where $G_H\ne\mathcal{C}$.
We call these \rk{phases} anomalous topological phases (ATP). There have been several numerical studies of \sat{topological systems} coupled to microscopic models of leads (called active leads) \cite{Jiang2009NumericalTAI, Bhat2025, Gusev2019MesoscopicTransport, AgarwalaShenoy2017TopologicalAmorphous, huang2018quantum}. None of these have reported any ATP.

A theoretical analysis by Buttiker, Pretre, and Thomas ~\cite{buttiker1994current,Brouwer1998}, argued that $G_H(\mu)$ as defined by charge pumping is equal to a boundary invariant given by the winding of the phase of the reflection matrix as a function of the flux being threaded through the cylinder. This has been verified \cite{Meidan2011,Fulga2012}
in microscopic models with effectively one-dimensional 
leads. These works also do not report any ATPs.
 
In this work, we consider a static cylindrical setup where a voltage bias is applied to the leads at the two ends of the CI and a Hall current is induced in the transverse direction. We consider two-dimensional leads with anisotropic hopping parameters.
We employ the microscopic approach~\cite{ref_datta1997,ryndyk2016,ref_dhar2006, Waintal2024ComputationalQuantumTransport,BhatDhar2021EquivalenceNEGFScatteringKitaev} of scattering theory and nonequilibrium Green's function (NEGF) to compute the Hall conductance of a CI connected to metallic leads  (see Fig.~\eqref{fig: Open system setup}).  Here we  summarize the most important results:   

\noi (i) We find that a Hall current is induced in the leads, which is surprising since the isolated lead is time-reversal symmetric.
The induced chiral current is due to the coupling with the CI, which breaks time-reversal symmetry. We call this a chiral proximity effect. We obtain the contributions to the Hall conductance from the leads and the CI, and show that their sum is quantized, $G_H=n$ (in units of $e^2/h$), consistent with
Laughlin's general argument.

\noi (ii) We identify $n$ with a boundary topological invariant given by the winding number of the phase of the reflection coefficient, along the Fermi surface of the scattering states. This is similar to the 
topological invariant obtained in the charge pumping  setup~\cite{Meidan2010,Meidan2011,Fulga2012}. 

\noi (iii) We show that the value of $G_H=n$ depends on the model parameters ($\mu,\eta_x,\eta_y,\eta_c,\mu_w$)
defined below, leading to a complex topological phase diagram in the parameter space. For some phases, $G_H=\mathcal{C}$ and in other phases, $G_H\neq\mathcal{C}$, which we call ATP.

\noi (iv) Surprisingly, we find ATP even in the weak coupling limit.
We provide a physical explanation of this as coming from the 
translational symmetry of our model in the transverse direction. 

We now present the important steps in the calculations. The detailed derivations are provided in the Supplemental Material (SM)~\cite{SM}.

\textit{The Model:} 
We consider the two-dimensional~(2D) spinless Bernevig-Hughes-Zhang~(SBHZ) model~\cite{ref_bernevig2006,qi2011} on a cylindrical lattice of size $N_x \times N_y$. Periodic boundary conditions are applied along the $y$-direction, and semi-infinite metallic leads are attached at the boundaries in the $x$ direction, as shown in Fig.~\ref{fig: Open system setup}(a). Thus, the Hamiltonian of the full system is given by $\mathcal{H}_{\rm total} = \mathcal{H}^{S} + \mathcal{H}^L + \mathcal{H}^R + \mathcal{H}^{SL} + \mathcal{H}^{SR},$ where $\mathcal{H}^S$ is the Hamiltonian of the SBHZ wire,
\begin{eqnarray} \label{eq: SBHZ hamiltonian}
&&\mathcal{H}^{S} 
    = \sum_{x=0}^{N_x-1} \sum_{y=0}^{N_y-1}  \mu_w \bm{c}^\dagger(x, y) \sigma_x \bm{c}(x, y) \nonumber \\
&&+\eta \sum_{x=0}^{N_x-2} \sum_{y=0}^{N_y-1} \left[ \bm{c}^\dagger(x, y) \frac{(\sigma_x - i \sigma_y)}{2} \bm{c}(x+1, y) + {\rm H.c.} \right]\nonumber\\
&&+\eta \sum_{x=0}^{N_x-1} \sum_{y=0}^{N_y-1} \left[ \bm{c}^\dagger(x, y) \frac{(\sigma_x - i \sigma_z)}{2} \bm{c}(x, y+1) + {\rm H.c.}\right],
\end{eqnarray}
where $\bm{c}(x, y) = \begin{pmatrix}c_1(x,y),c_2(x,y)\end{pmatrix}^T$, with $c_{1}(x,y)$ and $c_{2}(x,y)$ denoting the fermionic annihilation operators corresponding to the two orbital degrees of freedom at lattice site $(x,y)$. Henceforth, we work in units with $\eta=1$ and $e=h=1$. The matrices $\sigma_x$, $\sigma_y$, and $\sigma_z$ are the usual Pauli matrices. For $0<\mu_w<2$ and $-2<\mu_w<0$, the isolated SBHZ model on a torus exhibits topologically nontrivial phases with Chern numbers $\mathcal{C}=+1$ and $\mathcal{C}=-1$, respectively. For an isolated model on a cylinder in these phases, the bulk band gap hosts edge states that are localized along the system boundaries and carry dissipationless chiral currents~\cite{ref_bernevig2013,Shankar2018}.  The Hamiltonians of the left and right leads are denoted by $\mathcal{H}^L$ and $\mathcal{H}^R$ and are modeled by nearest-neighbor tight-binding 2D lattices with hopping amplitudes, $\eta_x$ and $\eta_y$, in the longitudinal and transverse directions, respectively. Throughout this paper, we work in the limit where the bandwidth of the reservoir is large compared to the bandgap of the insulator, i.e., $\eta_x>|\mu_w|$, and for simplicity we take $\eta_y>0$. The 
leads are connected to the boundary sites of the SBHZ model, as shown in Fig.~\ref{fig: Open system setup}(a). The contacts are described by tight-binding Hamiltonians with hopping amplitude $\eta_c$. Details of the lead and coupling Hamiltonians are provided in the SM~\cite{SM}. 

\begin{figure}
\centering
\includegraphics[width=0.9\linewidth]{Setup_shifted_origin.pdf}
\caption{(a) Schematic of a Chern insulator connected to metallic leads in the cylindrical geometry. (b) Equivalent Rice-Mele type one-dimensional setup obtained after Fourier transforming along the $y$-direction. The crossed sites indicate the sites belonging to the leads. The parameters $a(k_m) = \sin k_m$ and $\eta_m$ indicate onsite potentials in the CI and lead respectively, $b(k_m) = \mu_w + \cos k$ and $\eta_x$ are hopping terms in the CI and lead respectively, and $\eta_c$ is the CI-lead coupling. }
\label{fig: Open system setup}
\end{figure}

The cylindrical geometry, and translational invariance in the $y$-direction, allows us to reduce the problem to that of a set of decoupled one-dimensional chains, each labeled by $k_m=2\pi m/N_y$, $m=-N_y/2,-N_y/2+1,-N_y/2+2,..,N_y/2-1$.   Fig.~\ref{fig: Open system setup}(b) shows the hoppings and the onsite chemical potential for a single chain with $a_m=\sin k_m$ and $b_m=\mu_w+\cos k_m$.


The system is allowed to evolve from the initial state where the wire is in an arbitrary state, and the reservoirs are described by grand canonical ensembles with temperatures $T_L=T_R=0$ and Fermi levels $\mu_L=\mu+\Delta\mu, \mu_R=\mu$,  where $\mu$ is  taken to lie in the bulk gap of the CI. In the long-time limit, the system attains a non-equilibrium steady state (NESS), in which a net current $\Delta J_y(\mu)$ flows in the $y$-direction as the system breaks time-reversal symmetry. We are interested in the Hall conductance, defined as  $G_H(\mu)=\lim_{\Delta\mu\rightarrow 0}\frac{\Delta J_y(\mu)}{\Delta \mu}$, in the thermodynamic limit $N_x,N_y\rightarrow\infty$. The Hall conductance can be computed using two different but equivalent approaches: the scattering approach and the NEGF formalism. In the next section, we present the details of the scattering approach, and we leave the details of the NEGF approach to the Supplementary Material~\cite{SM}.

\textit{Scattering method:} In the NESS, the scattering states with electrons incoming from the left (left scattering states) are filled up to the Fermi level $\mu+\Delta\mu$, and the scattering states with electrons incoming from the right (right scattering states) are filled up to the Fermi level $\mu$. Thus, up to the Fermi level $\mu$, the current due to the left scattering states is canceled out by the current due to the right scattering states. As a result of this,  only the left scattering states with energies between $\mu$ and $\mu+\Delta\mu$ contribute to the current. Since $\mu$ lies within the bulk gap of the CI, such scattering states decay exponentially away from the left junction into the CI.  Therefore, in the limit $N_x\rightarrow\infty$, we can ignore the right junction for the solutions of the scattering states that decay inside the CI and just consider a system with a single lead-CI junction.

The wavefunction of the scattering state of energy $\epsilon(q,k)$, that decays into the CI, is of  the form $\zeta_{q,k}(x,y)=\Psi_{q,k}(x)e^{i k y}$, where 
 we define $\Psi_{q,k}(x)$ as follows,
\begin{align} \label{eq:CI ansatz}
    \Psi_{q,k}(x)= 
    \begin{cases}
       e^{iqx} + r\,e^{-iqx}, & x \le -1, \\[4pt]
        u\begin{pmatrix} 1 \\ f \end{pmatrix} \lambda_1^{x} & x \ge 0.
\end{cases}
\end{align}
where  $\lambda_1,f,r,u$ are the unknown quantities to be determined. Physically, the above form represents an electron plane wave with momentum $q\in(0,\pi)$ incident at the junction from the lead, which is reflected with a reflection coefficient $r$ and decays inside the CI. Thus we have $\vert\lambda_1\vert<1$. Since wavefunctions decay exponentially in the CI, the transmission coefficient is zero. Hence $\vert r(q,k)\vert=1$.

Following the standard scattering approach, we find the exact solutions for the unknown quantities as (see \cite{SM} for details): 
$\epsilon=2\eta_x\cos q+2\eta_y\cos k$, $\Delta\equiv\frac{\epsilon^2-\sin^2 k-(\mu_w+\cos k)^2-1}{\mu_w+\cos k}$, $\lambda_{1,2} = \frac{\Delta \pm \sqrt{\Delta^2 -4}}{2}$, and $f=\frac{\lambda_1 +\mu_w + \cos k}{\epsilon + \sin k}$.   $\lambda_1$ is chosen to be the solution with absolute value less than 1, and since $\lambda_1\lambda_2=1,$ the other solution, denoted by $\lambda_2$, has magnitude greater than 1. Note  that $\lambda_1$ and $f$ are independent of the lead coupling
$\eta_c$. The other two parameters do depend on $\eta_c$ and are given by
\begin{equation}\label{eq:ur}
u=\frac{2i\eta_x\eta_c\sin q}{\eta_xc+ \eta_c^2e^{iq}},~~
r=-\frac{\eta_x c+\eta_c^2e^{-iq}}{\eta_x c+\eta_c^2e^{iq}}\equiv e^{-i\phi(q,k)},
\end{equation}
where $c=-\lambda_1 f,$ $\phi(q,k)=2\arctan[(c+\Lambda)/\gamma],~\gamma=-(\eta_c^2/\eta_x)\sin q$ and  $\Lambda=(\eta_c^2/\eta_x) \cos q.$ 

In the linear response regime, the local Hall conductance $\tilde{G}_H(\mu,x)=\partial \tilde{J}_y(\mu,x)/\partial \mu$, where $\tilde{J}_y(x)$ is the local current in the $y$ direction. We then obtain~\cite{SM}
\begin{equation} \label{eq: local conductance1}
\tilde{G}_{H}(\mu, x)= \frac{1}{2\pi}\int_{-\pi}^{\pi} dk \int_0^{\pi} dq~j_y(x ; q,k)~\delta ( \mu - \epsilon_{q, k}),
\end{equation}
where  $j_y(x;q,k)$ is the single particle local $y$-current density (see Ref.~\cite{SM}).
The integration is carried out over the Fermi surface defined by $\mu=\epsilon(q,k)$. We parameterize the Fermi surface by $(q_F(k),k)$, $\mu={\epsilon(q_F(k),k)}$.
The range of $k$ depends on the parameters $\eta_x,~\eta_y$ and $\mu$, and we define this to be the set $\mathcal{I}\subseteq (-\pi,\pi)$. For $\mu\ge0$, we have three cases of ranges, corresponding to three qualitatively different shapes of the Fermi surface for $\mu>0$ \cite{SM}. The three cases for $\mathcal{I}$ are,
\begin{itemize}[nosep]
\item[(I)] $(-\pi,\pi), \text{ if }  \Delta\eta> 0,~\text{and}~\mu<2 \Delta \eta,$
\item[(II)] $(-k_2,k_2),  \text{ if } \mu\geq 2\Delta \eta,$
\item[(III)] $(k_1,k_2)\cup(-k_2,-k_1),  \text{ if } \Delta \eta <0,~\text{and}~\mu<2 \Delta \eta.$
\end{itemize}
 Here $\Delta \eta \equiv |\eta_x-\eta_y|, ~k_1=\arccos[(\mu+2\eta_x)/(2\eta_y)]$, and $k_2=\arccos[(\mu-2\eta_x)/(2\eta_y)].$    

 Using the scattering state wavefunction given in Eq.~\eqref{eq:CI ansatz}, we get the following expressions for the local Hall conductance inside the lead and the CI,
\begin{align}
\tilde G_H^{\text{lead}}(\mu,x) &= \frac{2\eta_y}{\pi} \int_{k\in\mathcal{I}} dk \frac{\sin k~\cos (2q_Fx + \phi(q_F,k))}{\sqrt{4\eta_x^2 - (\mu - 2\eta_y \cos k)^2}},\label{eq:GHlmux}\\
\tilde G_H^{\text{CI}}(\mu,x) &= \frac{1}{\pi}\int_{k\in\mathcal{I}}dk ~\gamma\lambda_1^{x}\frac{(1 - f^2)\cos k - 2f \sin k}{\gamma^2 + (c+\Lambda)^2}.\label{eq:GHcimux}
\end{align}
\begin{figure}[t]
\centering
\includegraphics[width=1\linewidth]{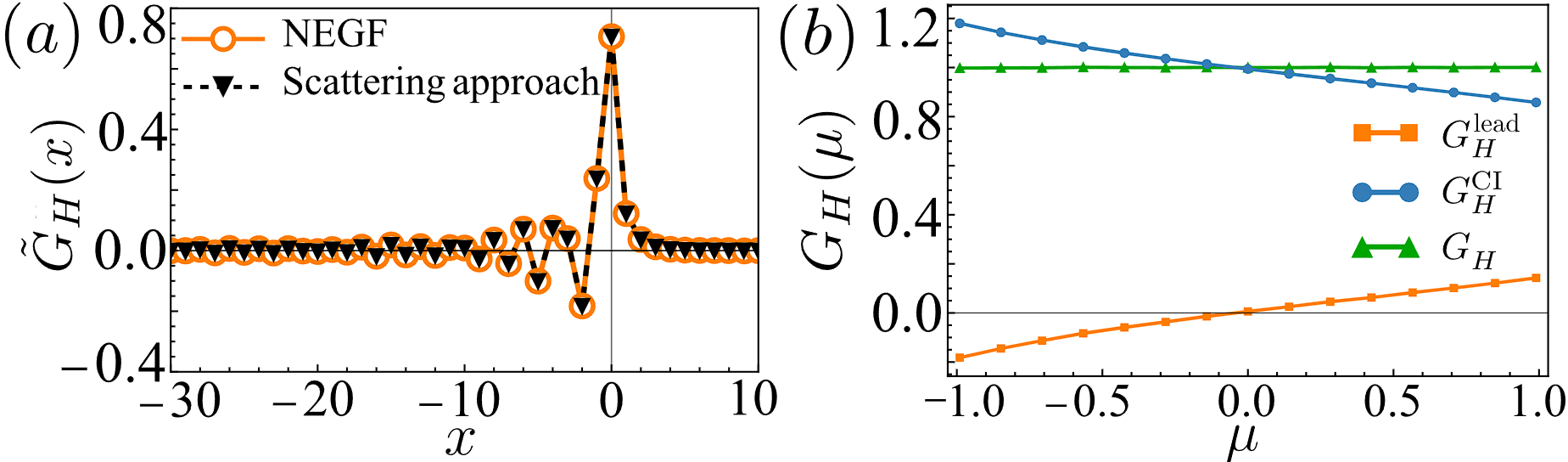}
\caption{(a) shows the local Hall conductance inside the leads and the CI at $\mu=0.8$. For a consistency check, the local Hall conductance was computed using two different methods, and we find perfect agreement. (b) shows contributions to the Hall conductance from the CI and metallic leads plotted as a function of the Fermi level $\mu$.  For both plots, $\mu_w= 1, \eta_x = 1, \eta_y = 0.25$, and $\eta_c = 1$.}
\label{Fig:Num_results_2Dleads}
\end{figure}

In Fig.~\ref{Fig:Num_results_2Dleads}(a), we plot $\tilde G_H(\mu,x)$ as a function of $x$ and see that it is non-vanishing in the leads -- we refer to this induced Hall conductance in the lead as a chiral proximity effect. Inside the leads,  we find that $\tilde{G}_H^{\text{lead}}(\mu,x)$ shows an interesting transition in its behavior with respect to $x$. For cases (I) and case (II) of $\mathcal{I}$,  we find a Friedel-type oscillatory decay as $\tilde G_H^{\text{lead}}(\mu,x)\sim \Re[Ae^{i2q_F(0)x}+Be^{i2q_F(\pi)x}]/|x|^{3/2}$ (for case (II), $B=0$ as $k=\pi \notin \mathcal{I}$)  while for case (III)  the oscillations disappear and $\tilde{G}_H^{\text{lead}}(\mu,x)$ decays exponentially with respect to $x$. For detailed explanations of these results, see Ref.~\cite{SM}. 

The total Hall conductance inside the leads and the CI is obtained by summing up the local Hall conductance over the respective regions, i.e., $G_H^{\rm lead}(\mu)=\sum_{x=-\infty}^{-1} \tilde{G}_H(\mu,x)$,  $G_H^{\rm CI}(\mu)=\sum_{x=0}^\infty \tilde{G}_H(\mu,x)$. The plot in Fig.~\ref{Fig:Num_results_2Dleads}(b)  shows the contributions of the lead and CI parts to the Hall conductance at different chemical potentials. Remarkably, we see that the sum of their contributions is quantized. Thus, the chiral proximity effect has to be taken into account to identify the Hall conductance with a topological invariant.

\textit{Hall conductance as a topological invariant:} The total Hall conductance is given by $G_H(\mu)=G_H^{\text{CI}}(\mu)+G_H^{\text{lead}}(\mu)$. Using Eqs.~\eqref{eq:GHlmux} and ~\eqref{eq:GHcimux}, we find after some algebra~\cite{SM} that the total Hall conductance can be expressed as
\begin{equation}
G_H(\mu)=\frac{1}{2\pi}\int_{k\in\mathcal{I}} dk\frac{d\phi(q_F(k),k)}{dk},\label{eq:wn}
\end{equation}
where  $\phi(q_F(k),k)$ is the phase of the reflection coefficient, $r$. Eq.~\eqref{eq:wn} implies that
$G_H$ is quantized to an integer for all three cases of the interval $\mathcal{I}$. For case (I), the Fermi surface is a loop in $(q,k)$ space, hence $G_H$ is a winding number and hence an integer. For cases (II, III), 
since  $r(0,k)=r(\pi,k)=-1,~\forall k$,  $r$ is the same at the endpoints of the interval $\mathcal{I}$, and thus the endpoints of $\mathcal{I}$ can be identified.  It then again follows from
Eq.~(\ref{eq:wn}) that $G_H$ is a winding number and hence an integer. Thus, we have shown that in our model, 
$G_H$ is always quantized to an integer $n$ when the
Fermi level is in the gap. This is consistent with
Laughlin's general argument.

We now present our surprising results 
that $n$ is {\emph{not}} independent
of the lead parameters. There is a complex topological phase diagram
defined by $n(\mu,\eta_x,\eta_y,\eta_c,\mu_w)$. Many of these
phases are anomalous, namely $n\ne{\mathcal C}$. We compute $n$ using the fact that it is the number of times $\phi(q_F,k)$ crosses the value 
$\pi$ from below, minus the number of times it does so from above, as we go around the Fermi surface by increasing $k$ over the allowed range. We determine these 
points on the Fermi surface using Eq.~(\ref{eq:ur}). Then, by analyzing $\phi(q_F,k)$ in the vicinity of these points it is easy to determine whether the point $\phi=\pi$ is being crossed from below or above.

\begin{figure}[t]
\centering
{\includegraphics[width=\linewidth]{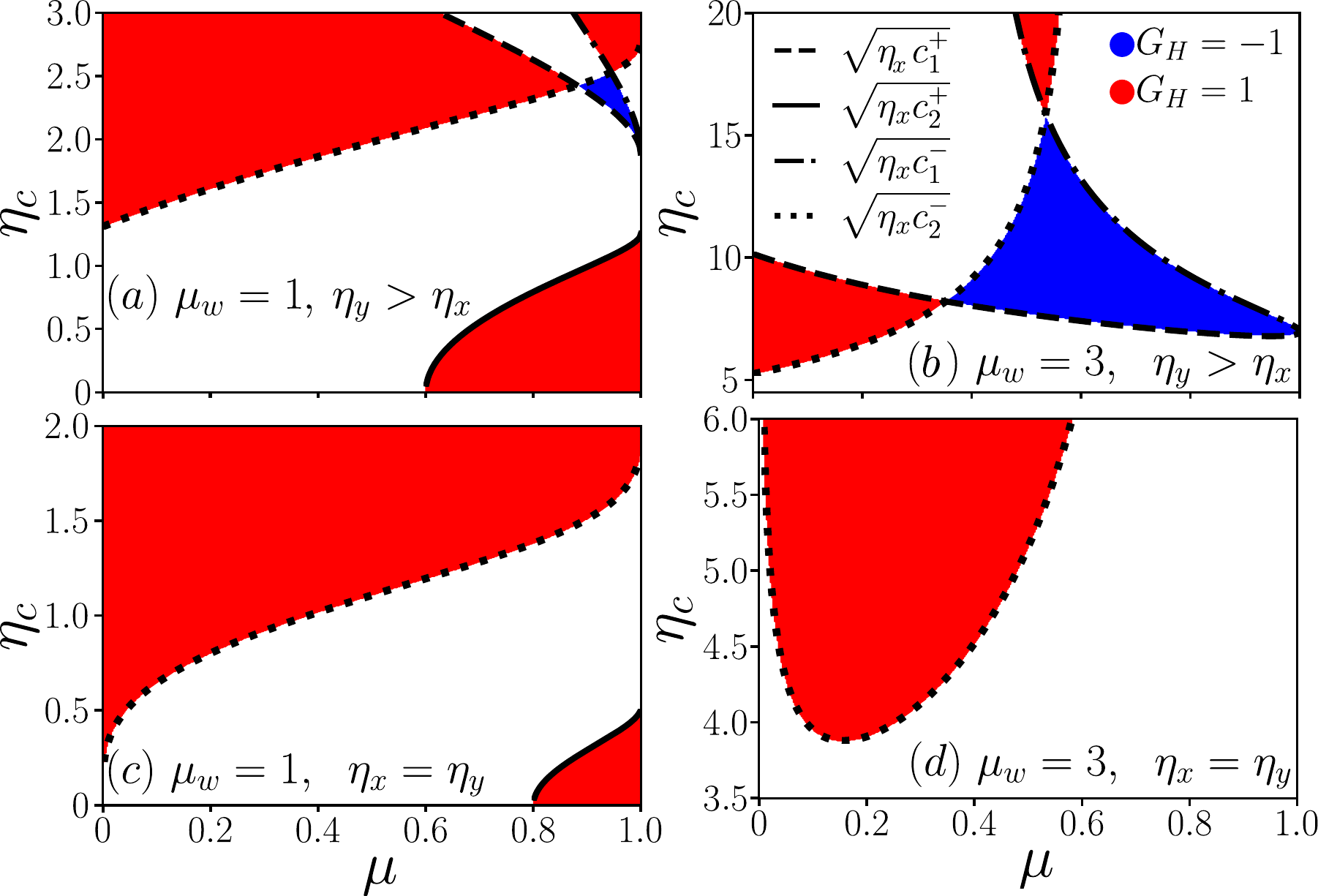}} 
\caption{Hall conductance for the regime  $(\mu_w=1,\mathcal{C}=1)$ and the  regime  $(\mu_w=3,\mathcal{C}=0)$ for different Fermi levels $\mu$ and coupling strengths $\eta_c$. (a) and (b) are for anisotropic leads corresponding to case (III)  leads $(\eta_y=4,~\eta_x=3.5)$, while (c) and (d) are for isotropic leads ($\eta_x=\eta_y=1$) that correspond to case (II). We have denoted $c_{1/2}^\pm=|c(\pm k_{1/2})|$. There are ATPs in all these cases. However, ATPs in the weak coupling limit occur only when $\mathcal{C}=1$.}
\label{fig:quant1}
\end{figure}

As detailed in the End Matter, applying this method to the case (I) when $\eta_x>\eta_y$ and $\mu<2\Delta\eta$,
we get $n=\mathcal{C}$ \cite{SM}. Thus in this regime of parameters we do not have any ATPs.
For cases (II,III), we find many ATPs, namely with $n\ne\mathcal{C}$. The calculations are detailed in the End Matter, and the results are shown in Fig.~\ref{fig:quant1}. Figs.~\ref{fig:quant1}(a) and ~\ref{fig:quant1}(c) show the phases corresponding to different quantized values of $G_H(\mu)$ in the $(\mu,\eta_c)$ space for $\mathcal{C}=1$, and Figs.~\ref{fig:quant1}(b) and ~\ref{fig:quant1}(d) show the phases for $\mathcal{C}=0$, respectively. We observe regions with $G_H=\pm 1$ in the $(\mu,\eta_c)$-space irrespective of whether $\mathcal{C}$ is zero or 1.

Surprisingly, we find ATPs even in the weak coupling limit. Namely $G_H(\mu)=0$
when $\mathcal{C}=1$ (the white regions in Figs.~\ref {fig:quant1}(a) and ~\ref{fig:quant1}(c)). 
As detailed in \cite{SM}, this is due to the translational symmetry in the $y$-direction. This imposes a kinematical constraint that prevents the topological edge states of the isolated CI from hybridizing with the incoming plane waves from the lead. 

ATPs also occur at strong couplings in Fig.~\ref{fig:quant1}(a) where we see that $G_H(\mu)=-1$ even if $\mathcal{C}=1$, and in Fig.~\ref{fig:quant1}(b), Fig.~\ref{fig:quant1}(d) where we see regions of $G_H(\mu)=\pm 1$ even if $\mathcal{C}=0.$ We do not have a physical interpretation of this so far. 

From Fig.~\ref{fig:quant1}, we see that the phase boundaries are given by the curves $\eta_c^2=\eta_x |c(\pm k_{1/2})|$ (see End Matter for details). Since the system is gapless, there is no concept of a gap closing at these boundaries. However,  the reflection coefficient and hence its phase are undefined at least at one point on the Fermi
surface \cite{SM} at the phase boundaries. To see this, we consider the point $k=k_1$ of the Fermi surface where $q_F(k_1)=\pi$, and we find that $r(k_1)=-\frac{\eta_x c(k_1) -\eta_c^2}{\eta_x c(k_1)-\eta_c^2},$ which is clearly undefined if $\eta_c^2=\eta_x c(k_1)$, assuming that $c(k_1)>0$. 

\textit{Discussion:} We have shown that the model we 
have analyzed has a complex topological phase diagram 
in its 
five-parameter space as illustrated in Fig.~\ref{fig:quant1}.
A striking feature is that we get ATPs even in the weak
coupling limit $\eta_c\rightarrow 0$, as illustrated
in the figure. This is due to the translational symmetry in the $y$ direction of our model, which implies that the edge states of the isolated CI
can only hybridize with the plane waves of the leads when their energies {\em and} their wave vectors in the $y$-direction are equal. These kinematical constraints prevent the edge channel 
states of the isolated CI from hybridizing with the plane
waves of the leads for some ranges of the lead parameters \cite{SM}.
Consequently, in such cases, the leads are unable to probe
the BBC, leading to ATPs in the weak coupling limit. Furthermore, we also find ATPs in the strong coupling limit whose physical origin is not clear.

A natural question to ask is whether the above physics is sensitive to the microscopic details of the coupling between the CI and the leads. We have checked numerically that our results are true for a different lead-CI coupling Hamiltonian, one where both degrees of freedom inside the CI are coupled to the leads \cite{SM}. 
 
While there are heuristic arguments that suggest that our definition of $G_H$ exactly the same as the definition 
using the charge pumping \cite{laughlin1981quantized}, a proof of this in a microscopic model is desirable. It may 
be possible to experimentally verify our predictions in ultra-cold atom 
systems~\cite{celi2014synthetic,boada2012extra,Han2019SyntheticHallTube,Li2022BEC,fabre2022laughlin}. The Corbino geometry may also provide another  possibility to experimentally test our predictions, since the rotational symmetry of the system may impose the same
kinds of kinematic constraints as the translational symmetry of our system in the cylindrical geometry.

\section*{Acknowledgements}
We thank Bijay Kumar Agarwalla, Ankush Chaubey, Beno\^{i}t Dou\c{c}ot, Abhinav Dhawan, Joel Moore, Ganapathy Murthy, Subir Sachdev, Ashoke Sen and T. Senthil for very helpful discussions. SS acknowledges the LTVS Program for funding the visit to ICTS. AD acknowledges the J.C. Bose
Fellowship (JCB/2022/000014) of the Science and Engineering Research Board of the Department of Science and Technology, Government of India. SS, RK and AD	acknowledge the support of the DAE, Government of India,
	under projects nos. 12-R\&D-TFR-5.10-1100 and RTI4001.

\bibliography{new_refs_bib.bib}
\newpage

\begin{widetext}

\begin{center}
    \section{End Matter}
\end{center}
\label{sec:endmatter}
\vskip .2cm
We present the algorithm for the computation of the boundary topological invariant for different parameter regimes. We focus on  $\mu>0$, lying inside the bulk gap of the CI, and $\mu_w>0$. This is the algorithm 
implemented to produce Fig \ref{fig:quant1}.

As stated in the main text, we compute the winding number of
$\phi(q_F(k),k)$ by determining the number of times it crosses $\pi$ from below minus the number of times it does so from above. From Eq.~\eqref{eq:ur}, $\phi(q_F(k),k)=\pi$ implies that $q_F(k)=0,\pi$ or
$c(k)=\pm\infty$, or both. We can express $c(k)$ as
\begin{equation}
\label{cdef}
c(k)=\frac{\sin k-\mu}
{1+\lambda_1(q_F(k),k)(\mu_w+\cos k)},
\end{equation}
where $\vert\lambda_1(q_F(k),k)\vert<1$. Thus a necessary condition
for $c(k)=\pm\infty$ is that the denominator of $c(k)$ vanishes which gives
\begin{equation}
\label{nc1}
\lambda_1=-\frac{1}{\mu_w+\cos k}~\Rightarrow
\vert \mu_w+\cos k\vert >1.
\end{equation}
After straightforward algebra, we find that $\lambda_1=-1/(\mu_w+\cos k)$ implies 
$\sin k=\pm \mu.$

\begin{enumerate}
\item
When $\sin k=-\mu$, Eq.~\eqref{cdef} implies that
$c(q_F(k),k)=\pm\infty$ provided $|\mu_w+\cos k|>1$.
\item
When $\sin k=\mu$ the application of the
L'Hopital's rule shows that $c(k)$ is finite.
\end{enumerate}
Thus when $\mu>0$, $c(k)$ can diverge only when
$k$ is in the third or fourth quadrant. We denote the solutions of $\sin k=-\mu$ in the third and fourth quadrants by $k^*_3=-\pi+\arcsin\mu$ and $k^*_4=-\arcsin\mu$, respectively. We now consider our three cases separately.

\subsection{Case I}
\begin{figure}[t]
\centering
\includegraphics[width=0.8\linewidth]{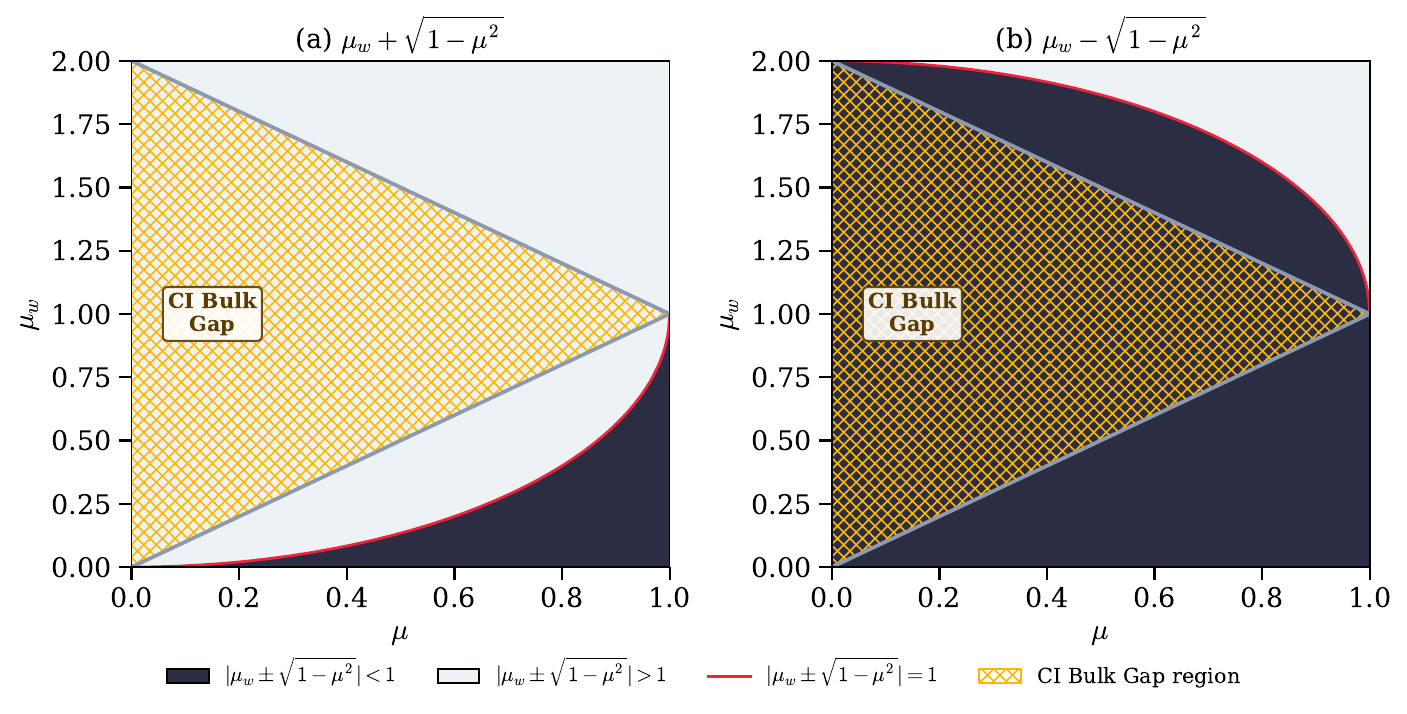}
\caption{(a) and (b) show the regions where the functions $\mu_w\pm\sqrt{1-\mu^2}$ are greater than or less than 1. It can be clearly seen that as $\mu$ lies in the bulk gap of the CI, $|\mu_w-\sqrt{1-\mu^2}|<1$ and $|\mu_w+\sqrt{1-\mu^2}|>1$ for $\mu_w\in(0,2)$.}
\label{fig:muwmu}
\end{figure}

In this case, $\mathcal{I}\in(-\pi,\pi)$. Further, $q_F(k)\ne 0,\pi$. Thus $\phi(q_F(k),k)$
crosses $\pi$ if and only if $c(q_F(k),k)=\pm\infty$. The number of times
that $\phi$ crosses $\pi$ from below minus the number of
times it does so from above can be computed by the 
following algorithm.
\begin{enumerate}
\item 
When $\mu_w>2$ ($\mathcal{C}=0$), then $\vert \mu_w+\cos k^*_{3,4}\vert >1.$ Thus $\phi$ crosses $\pi$ at $k^*_3$ and $k^*_4.$ Expanding  $c(k)$ around $k^*_3$ and $k^*_4$ gives the leading behavior to be,
\begin{equation}
    c(k)\approx \frac{(\mu_w+\cos k_{3,4}^*)^2-1}{\cos k_{3,4}^*}\frac{1}{k-k_{3,4}^*}.
    \label{aeq:ckexp}
\end{equation}
Since $\cos k_3^*<0$ and $\cos k_4^*>0$,  $\phi(k)$ jumps from $-\pi$ to $\pi$ at $k^*_3$, and then jumps from $\pi$ to $-\pi$ at $k^*_4$. Thus, $\phi(k)$ crosses $\pi$ in opposite directions, which implies that it does not wind and thus $G_H(\mu)=0$. 
\item 
When $0<\mu_w<2$, ($\mathcal{C}=1$), we need the values of  $b_3=|\mu_w+\cos k^*_3|$ and $b_4=\vert\mu_w+ \cos k^*_4\vert$ relative to 1 to determine whether $c(k_{3,4}^*)=\pm \infty$. We note that these two quantities can be rewritten as $b_3=|\mu_w-\sqrt{1-\mu^2}|$, $b_4=|\mu_w+\sqrt{1-\mu^2}|$, and Fig.~\ref{fig:muwmu} shows that $b_3<1$ and $b_4>1$ for all $\mu_w\in (0,2)$ as long as $\mu$ lies inside the bulk gap of the CI, i.e., $\mu<\min\{\mu_w,2-\mu_w\}$.   Thus $\phi$ crosses $\pi$ only at $k^*_4$. The expansion of $c(k)$ around $k^*_4$ in Eq.~\eqref{aeq:ckexp} shows that $\phi(k)$ jumps from $\pi$ to $-\pi$ at $k_4^*$ and therefore $G_H=1=\mathcal{C}$.
\end{enumerate}
Thus in this case, we have $G_H=\mathcal{C}.$

\subsection{Case II}
In this case we have $\mathcal{I}=(-k_2,k_2)$. The Fermi surface consists of a single loop
from $(0,-k_2)$ to ($0,k_2)$. Further, at the point $(0,\pm k_2)$, $\phi(0,\pm k_2)=\pi$. The number of times 
that $\phi$ crosses $\pi$ from below to above minus the number of times it does so from above to below at this 
point, $n'=\pm 1$, is given by
\begin{equation}\label{eq:n'}
n'=- \frac{1}{2}\left(\text{sgn}(c(k_2)-\eta_c^2/\eta_x)
-\text{sgn}(c(-k_2)-\eta_c^2/\eta_x)\right).
\end{equation}
Hence $n'$ is easily computable. There are other points on the Fermi surface where $\phi$ crosses $\pi$, namely points on the
Fermi surface where $c(q_F (k), k) =\pm\infty$. The algorithm to  compute  the number of times that $\phi$ crosses $\pi$ from
below and above in these cases, $n''$, has been discussed in Case I. Thus the winding number $n=n'+n''=G_H$ is easily computable.

\subsection{Case III}
In this case we have $\mathcal{I}=(k_1,k_2)\cup(-k_2,-k_1)$. Hence the Fermi surface consists of two loops, one 
going from $k_1$ to $k_2$ and the other going from $-k_2$ to $-k_1$. Further, $q_F(\pm k_1)=\pi$ and $q_F(\pm k_2)=0$.
It then follows from Eq. (\ref{eq:ur}) that $\phi(\pm k_{1,2})=\pi$. The number of times that $\phi$ crosses $\pi$ from below to above minus the number of times it does so from above to below at these points, $n'$,
is given by
\begin{equation}\label{eq:n'1}
n'=- \frac{1}{2}\left(\text{sgn}(c(k_2)-\eta_c^2/\eta_x)
-\text{sgn}(c(k_1)-\eta_c^2/\eta_x)+\text{sgn}(c(-k_1)-\eta_c^2/\eta_x)-\text{sgn}(c(-k_2)-\eta_c^2/\eta_x)\right).
\end{equation}
$n'$ is again easily computable. There are other points on the Fermi surface where $\phi$ crosses $\pi$, namely 
points on the Fermi surface where $c(q_F(k),k)=\pm\infty$. The methodology of computing the number of times
that $\phi$ crosses $\pi$ from below and 
above has been discussed in Case I. Thus, putting all the results together, the method for computing $G_H$ in this 
case is also clear.
\end{widetext}


\newpage
\onecolumngrid
\clearpage
\twocolumngrid

\appendix
\setcounter{section}{0}
\renewcommand{\thesection}{\Alph{section}}

\setcounter{figure}{0}
\renewcommand{\thefigure}{A\arabic{figure}}
\onecolumngrid

\begin{center}
	{\large\bfseries
		Supplementary Material: Anomalous topological phases in a Chern insulator connected to leads in the cylindrical geometry
	}
\end{center}
\twocolumngrid

\section{Glossary of symbols} \label{section: Glossary}

\begin{tabular}{|c|c|}
	\hline
	Symbol& Definition\\
	\hline
	$\mu_w$&CI parameter\\
	\hline
	$\eta_x$& $x$-direction hopping strength in the leads\\
	\hline
	$\eta_y$& $y$-direction hopping strength in the leads\\
	\hline
	$\eta_c$&Coupling of leads to CI\\
	\hline
	$\epsilon(q,k)$& $2(\eta_x\cos q+\eta_y\cos k)$\\
	\hline
	$\Lambda(\omega,k)$&$\frac{\eta_c^2}{2\eta_x^2} (\omega-2 \eta_y \cos k)$\\
	\hline
	$\gamma(\omega,k)$&$-\frac{\eta_c^2}{2\eta_x^2} [4 \eta_x^2-(\omega-2 \eta_y \cos k)^2]^{1/2}$\\
	\hline
	$a(k)$& $\sin k$\\
	\hline
	$b(k)$&$\mu_w+\cos k$\\
	\hline
	$\Delta(q,k)$&$\frac{1}{b}\left(\epsilon^2-a^2-b^2-1\right)$\\
	\hline
	$\lambda(q,k)$&$\frac{1}{2}\left(\Delta\pm\sqrt{\Delta^2-4}\right)$\\
	\hline
	$\lambda_1(q,k)$&$ \lambda$ if $\vert\lambda\vert<1$\\
	\hline
	$\lambda_2(q,k)$& $\lambda$ if $\vert\lambda\vert >1$\\
	\hline
	$f(\mu,k)$&$(\epsilon-a)/(\lambda_2+b)$\\
	\hline
	$g(\mu,k)$&$(a-\epsilon)+\Lambda+bf$\\
	\hline
	$c(q,k)$&$(a-\epsilon)+bf$\\
	\hline
	$h(k)$&$((1 - f^2)\cos k - 2af)/(1-\lambda_1^2)$\\
	\hline
	$\phi(\mu,k)$&$2\tan^{-1}\frac{c + \Lambda}{\gamma}$\\
	\hline
\end{tabular}
\section{Setup details and transformation to 1D chains}
\label{section: Setup details}
\begin{figure}
	\centering
	\includegraphics[width=.9\linewidth]{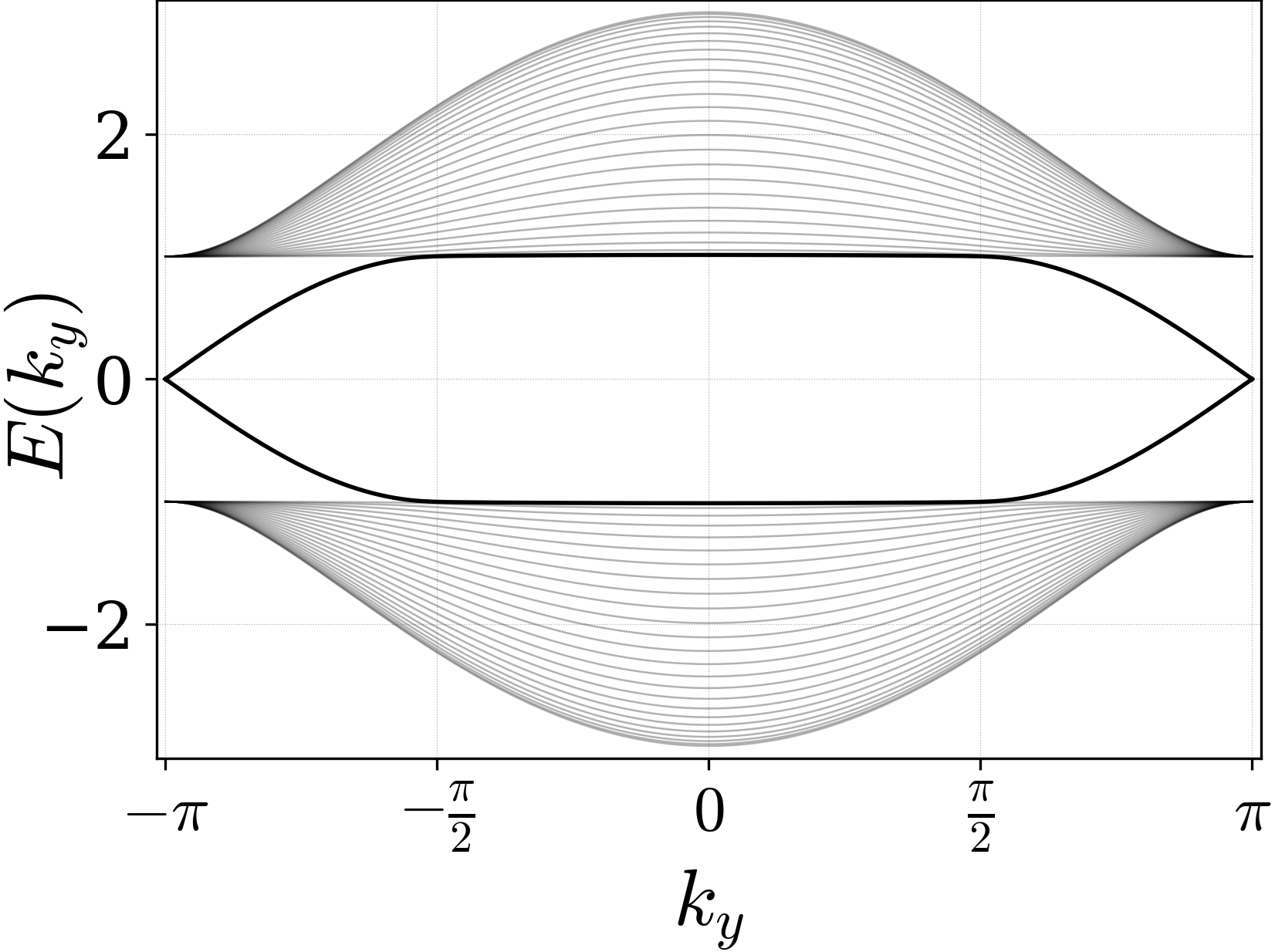}
	\caption{\textbf{Band diagram:} Plot of the spectrum of the isolated CI for $\mu_w=1$. The solid black lines inside the band gap correspond to edge states localized on the left and right edge of the cylinder. The edge state eigenvalues are given by $\epsilon_k=\pm \sin k$ where $k$ satisfies the condition $|\mu_w+\cos k| <1$.}
	\label{fig:CIspectrum}
\end{figure}

As mentioned in the main text, the Hamiltonian for the full system is given by :
\begin{equation}
	\mathcal{H}_{\rm total} = \mathcal{H}^{S} + \mathcal{H}^L + \mathcal{H}^R + \mathcal{H}^{SL} + \mathcal{H}^{SR}.
\end{equation}
We recall that $\mathcal{H}^{S}$ represents the Hamiltonian for the Chern insulator (CI) in cylindrical geometry, given by  
\begin{eqnarray} \label{eq: SBHZ hamiltonian1}
	&&\mathcal{H}^{S} 
	= \sum_{x=0}^{N_x-1} \sum_{y=0}^{N_y-1}  \mu_w \bm{c}^\dagger(x, y) \sigma_x \bm{c}(x, y) \nonumber \\
	&&+ \sum_{x=0}^{N_x-2} \sum_{y=0}^{N_y-1} \left[ \bm{c}^\dagger(x, y) \frac{(\sigma_x - i \sigma_y)}{2} \bm{c}(x+1, y) + h.c. \right],\nonumber\\
	&&+\sum_{x=0}^{N_x-1} \sum_{y=0}^{N_y-1} \left[ \bm{c}^\dagger(x, y) \frac{(\sigma_x - i \sigma_z)}{2} \bm{c}(x, y+1)+h.c. \right].
\end{eqnarray}
where $\bm{c}(x,y)=\begin{pmatrix}c_1(x,y)&&c_2(x,y)\end{pmatrix}^T$ and $c_{1}(x,y)$ and $c_{2}(x,y)$ are the fermionic annihilation operators corresponding to the two orbital degrees of freedom at lattice site $(x,y)$ of the CI. Note that we have set the hopping within the CI in Eq.~\eqref{eq: SBHZ hamiltonian1} to be $1$. The dispersion for the bulk states of the Hamiltonian $\mathcal{H}^S$ is given by
\begin{equation} \label{eq:CIdis}
	E(\mathbf{k})=\pm\sqrt{
		(\mu_w+\cos k_x+\cos k_y)^2
		+\sin^2k_x+\sin^2k_y},
\end{equation}
and the band diagram for this model is shown in Fig.~\ref{fig:CIspectrum}.
Note that in Eq.~\eqref{eq:CIdis}, $\mathbf{k} = (k_x, k_y)$ is the crystal momentum of the electrons and $\mu_w$  determines whether the isolated system is in a topologically trivial or nontrivial phase.

$\mathcal{H}^L$ and $\mathcal{H}^R$ represent the Hamiltonian of the left and right leads, respectively. $\mathcal{H}^{SL}\,(\mathcal{H}^{SR})$ represent the contacts between CI and the left (right) lead. The leads are modeled by the nearest-neighbor tight-binding Hamiltonians as follows,
\begin{eqnarray}
	\label{eq:H-L}
	\mathcal{H}^L &=& \sum_{x=-\infty}^{-2}\sum_{y=0}^{N_y-1}\eta_x \bigl[c_L^\dagger(x,y)\,c_L(x+1,y) + \text{h.c.}\bigr]\notag\\
	&+&\sum_{x=-\infty}^{-1}\sum_{y=0}^{N_y-1}\eta_y \bigl[c_L^\dagger(x,y)\,c_L(x,y+1) + \text{h.c.}\bigr],
	\label{eq:H-R}\\
	\mathcal{H}^R &=& \sum_{x=N_x}^{\infty}\sum_{y=0}^{N_y-1}\eta_x \bigl[c_R^\dagger(x,y)c_R(x+1,y) + \text{h.c.}\bigr]\notag\\
	&+&\sum_{x=N_x}^{\infty}\sum_{y=0}^{N_y-1}\eta_y \bigl[c_L^\dagger(x,y)\,c_L(x,y+1) + \text{h.c.}\bigr],
	\label{eq:H-R1}
\end{eqnarray}
where $c_{L}(x,y)$ and $c_{R}(x,y)$ are fermionic annihilation operators in the left and right leads, respectively, and  $\eta_x$ and $\eta_y$ are the hopping amplitudes along the $x$ and $y$ direction, respectively. We have imposed periodic boundary conditions along the $y$-direction throughout the setup, therefore $\mathbf{c}(x,N_y)=\mathbf{c}(x,0)$ and $c_{L/R}(x,N_y)=c_{L/R}(x,0)$ for all $x$. The contacts between CI and the leads are modelled by the following Hamiltonians,
\begin{eqnarray}
	\mathcal{H}^{SL} &=& \sum_{y=0}^{N_y-1}\eta_c\,\bigl[c_L^\dagger(-1,y)\,c_1(0,y) + \text{h.c.}\bigr],\\[1mm]
	\mathcal{H}^{SR} &=& \sum_{y=0}^{N_y-1}\eta_c\,\bigl[c_2^\dagger(N_x -1,y)\,c_R(N_x,y) + \text{h.c.}\bigr].\label{eq: Coupling Hamiltonian}
\end{eqnarray}
Here, $\eta_c$ is the coupling between the CI and the leads. Due to the translational symmetry along $y$-direction, $\mathcal{H}_{total}$ is block diagonal in the Fourier basis,
\begin{align}
	\bm{\tilde c}_m(x)&=\frac{1}{\sqrt{N_y}}\sum_{y=0}^{N_y-1} e^{-i k_m y}\bm{c}(x,y), \label{eq:Fourier_transform_in_ydirection}\\
	\tilde c_{L,m}(x)&=\frac{1}{\sqrt{N_y}}\sum_{y=0}^{N_y-1} e^{-i k_m y}c_L(x,y),\\
	\tilde c_{R,m}(x)&=\frac{1}{\sqrt{N_y}}\sum_{y=0}^{N_y-1} e^{-i k_m y}c_R(x,y),
\end{align}
where  $k_m=2\pi m/N_y$ with $m=-{N_y}/{2},-{N_y}/{2}+1...,{N_y}/{2} -1$ where we assume $N_y$ to be even. Using these operators we rewrite $\mathcal{H}_{\rm total}$ as follows,
\begin{align} \label{eq:Htotal_FT}
	\mathcal{\Tilde{H}}_{\rm total} =\sum_{m=-{N_y}/{2}}^{N_y/2 -1} \tilde{\mathcal{H}}^{S}_m + \tilde{\mathcal{H}}^L_m + \tilde{\mathcal{H}}^R_m + \tilde{\mathcal{H}}^{SL}_m + \tilde{\mathcal{H}}^{SR}_m,
\end{align}
where
\begin{eqnarray} \label{eq:systemH}
	\tilde{\mathcal{H}}^{S}_m &=&\sum_{x',x=0}^{N_x-1} \bm{\tilde c}_m^\dagger(x) \tilde{H}^S_m[x,x']\bm{\tilde c}_m(x').
\end{eqnarray}
Here, $\tilde{\mathcal{H}}^{S}_m$ denotes the many-body Hamiltonian of CI, whereas $\Tilde{H}^S_m[x,x']$ is a $2\times 2$ block of the single particle Hamilonian $\tilde{H}^S_m$ which has the following matrix elements,
\begin{equation}\label{eq: CImatrix}
	\Tilde{H}^S_m= \begin{pmatrix}
		V_m  &  A & 0 & 0 & \cdots & 0 & 0 \\
		A^{T} & V_m & A & 0 & \cdots & 0 & 0 \\
		0 & A^{T} & V_m &  A & \cdots & 0 & 0 \\
		\vdots & \vdots &  A^{T} & V_m & \ddots & \vdots & \vdots \\
		\vdots & \vdots & \vdots & \ddots & \ddots & A & 0 \\
		\vdots & \vdots & \vdots & \cdots & A^{T} &  V_m & A \\
		0 & 0 & 0 & \cdots & 0 &  A^{T} & V_m\\
	\end{pmatrix},
\end{equation}
where $V_m = a(k_m)\sigma_z + b(k_m) \sigma_x$ and $A = (\sigma_x - i\sigma_y)/2$ with $a(k_m)=\sin k_m$ and~$b(k_m)=\mu_w + \cos k_m$.  For the lead and coupling Hamiltonians, we have
\begin{align}
	\Tilde{\mathcal{H}}_m^L =& 2\eta_y\cos k_m \sum_{x=-\infty}^{-1}\Tilde{c}_{L,m}^\dagger(x)\,\Tilde{c}_{L,m}(x) \notag \\
	&+\sum_{x=-\infty}^{-2}\eta_x 
	\bigl[\Tilde{c}_{L,m}^\dagger(x)\,\Tilde{c}_{L,m}(x+1) + \text{h.c.}\bigr], \label{eq:HL_m)}
	\\ 
	\Tilde{\mathcal{H}}_m^R=&
	2\eta_y\cos k_m \sum_{x=N_x}^\infty\Tilde{c}_{R,m}^\dagger(x)\,\Tilde{c}_{R,m}(x) \notag \\& +
	\sum_{x= N_x}^{\infty} \eta_x \bigl[\Tilde{c}_{R,m}^\dagger(x)\,\Tilde{c}_{R,m}(x+1) + \text{h.c.}\bigr],\label{eq:HR_m)}
\end{align}
\begin{align}
	&\Tilde{\mathcal{H}}_m^{SL} = \eta_c\,\bigl[\tilde c_{L,m}^\dagger(-1)\,\tilde c_{1,m}(0) + \text{h.c.}\bigr], \label{eq: H_SL}\\
	&\Tilde{\mathcal{H}}_m^{SR} = \eta_c\,\bigl[\tilde c_{2,m}^\dagger(N_x-1)\,\tilde c_{R,m}(N_x) + \text{h.c.}\bigr].
\end{align}
Thus, taking the partial Fourier transform of $\mathcal{H}_S$ along the $y$-direction yields $N_y$ decoupled 1D chains, with Hamiltonians $\Tilde{H}^S_m$ resembling the Rice–Mele model~\cite{RiceMele1982,asboth2016short,dwivedi2016}. See Fig.~1 of the main text for a diagram of the setup considered in this work and its equivalent 1D setup obtained via partial Fourier transform.

\section{Hall conductance via scattering approach}\label{section: hall conductance via scattering approach}

In this section, we obtain the exact scattering state solution for a lead-CI system, and use it to compute the local Hall conductance given in Eq.~(4) of the main text. 
\subsection{Exact scattering state solution}\label{subsection:scattering state}
We consider a semi-infinite CI connected to a semi-infinite 2D metallic lead with periodic boundary conditions (PBC) along the 
$y$-direction. Using Eq.~\eqref{eq:systemH}, \eqref{eq:HL_m)}, \eqref{eq: H_SL} and replacing $k_m \rightarrow k$ in the limit $N_y \rightarrow \infty$, the Hamiltonian for this \emph{single} lead-CI system  is given by,
\begin{equation} \label{eq: H_LCI}
	\mathcal{\tilde{H}} = \int_{-\pi}^{\pi} dk~[ \tilde{\mathcal{H}}^{S}_k + \tilde{\mathcal{H}}^L_k + \tilde{\mathcal{H}}^{SL}_k]
	= \int_{-\pi}^{\pi} dk~\mathbf{C}_k^{\dagger}\tilde{H}^{\rm total}_k\mathbf{C}_k,
\end{equation}
where $\tilde{H}^{\rm total}_k$ represents the single-particle Hamiltonian for the single lead-CI setup and $\mathbf{C}_k$ represents the column vector formed using the annihilation operators at all lattice sites. Note that, here the annihilation operators are arranged in the same order as their corresponding lattice sites are labelled in Eq.~\eqref{eq:systemH}, \eqref{eq:HL_m)} and \eqref{eq: H_SL}.\\[2mm]
Due to the translational symmetry along the $y$-direction, the wavefunction of the scattering state is of the form $\zeta_{q,k}(x,y)=\Psi_{q,k}(x)e^{i k y}$ where for notational convenience, we define $\Psi_{q,k}(x)$ as,
\begin{align} \label{eq:CI ansatz1}
	\Psi_{q,k}(x)= 
	\begin{cases}
		\psi_M(x) , & x \le -1, \\[4pt]
		\begin{pmatrix}
			\psi_1(x)\\ \psi_2(x)
		\end{pmatrix} & x > -1.
	\end{cases}
\end{align}
$\psi_M(x)$ is the wavefunction inside the leads and $\begin{pmatrix}
	\psi_1(x), \psi_2(x)
\end{pmatrix}^T$ is the wavefunction inside the CI. Following the standard scattering approach, we first consider the eigenvalue equation 
\begin{equation}\label{eq:evaleqn}
	\tilde{H}^{\rm total}_k~\Psi_{q,k}=\epsilon\Psi_{q,k},
\end{equation}
in the bulk of the lead ($x<-1$) and the CI $(x>0)$. For the region inside the leads ($x<-1$), Eq.~\eqref{eq:evaleqn} gives
\begin{align}
	\eta_x\psi_M(x+1) + 2\eta_y \cos k\psi_M(x) + \eta_x \psi_M(x-1)  &= \epsilon \psi_M(x),\label{eq:metal_bulk_eqn}
\end{align}
and for the region inside the CI  Eq.~\eqref{eq:evaleqn} gives
\begin{align}
	\psi_2(x-1) + b\psi_2(x) + a\psi_1(x) &= \epsilon \psi_1(x),~~~ x \geq 1, \label{eq:CI_bulk_eqn1}\\
	\psi_1(x+1) + b\psi_1(x) - a\psi_2(x) &= \epsilon \psi_2(x),~~~ x \geq 0.\label{eq:CI_bulk_eqn2} 
\end{align}
We look for solutions of the  following form:
\begin{align}
	\label{eq:leadan} \psi_M(x) &= e^{iqx} + r\,e^{-iqx},~~x \le -1, \\
	\label{eq:CIan}  \begin{pmatrix}
		\psi_1(x)\\ \psi_2(x)
	\end{pmatrix} &=\begin{pmatrix} u \\ v \end{pmatrix} \lambda^{x},~~x > -1,
\end{align}
where $u,v,r,\lambda$ are unknown coefficients to be determined. Substituting Eq.~\eqref{eq:leadan} and Eq.~\eqref{eq:CIan} into Eq.~\eqref{eq:metal_bulk_eqn},~\eqref{eq:CI_bulk_eqn1},~\eqref{eq:CI_bulk_eqn2}, we get the dispersion as,
\begin{equation} \label{eq:dispersion}
	\epsilon = 2\eta_x \cos q + 2\eta_y \cos k,
\end{equation}
and the two possible values of $\lambda$ is given by
\begin{equation} \label{eq: lambda+-}
	\lambda_{1,2} = \frac{\Delta \pm \sqrt{\Delta^2 -4}}{2}, 
\end{equation}
where $\Delta=\frac{\epsilon^2 - a^2 - b^2 -1}{b}$. Here $\lambda_1$ and $\lambda_2$ are chosen such that  $|\lambda_1|<|\lambda_2|$.  We note that $\lambda_1 \lambda_2=1$, hence $|\lambda_1|<1$. We want an exponentially decaying solution inside the CI, and so we choose $\lambda = \lambda_1$. Finally, using bulk equations of CI (Eq.~\eqref{eq:CI_bulk_eqn1} and Eq.~\eqref{eq:CI_bulk_eqn2}) we also get
\begin{equation} \label{eq: u/v}
	\frac{v}{u}=f(k) = \frac{\epsilon -a}{b+ 1/\lambda_1} = \frac{\lambda_1 + b}{\epsilon + a}.
\end{equation}
Note that from Eq.~\eqref{eq: lambda+-}, the decaying solution inside the CI exists only for $|\Delta(k)|>2$, so it defines the band gap region. Whereas for $|\Delta(k)|<2$, we get extended solution; this corresponds to the bulk bands of the CI. Finally, the band edges is defined by the condition $|\Delta(k)|=2$.

The remaining unknowns now are the coefficients $u$ and $r$. These are fixed by the eigenvalue equation at the lead-CI junction given by,
\begin{align}
	\eta_x \psi_M(-2) + 2\eta_y \cos k\psi_M(-1) + \eta_c\psi_1(0)  &= \epsilon \psi_M(-1), \label{eq:boundary_eqn1}\\
	\eta_c \psi_M(-1) + a\psi_1(0) + b\psi_2(0) &= \epsilon \psi_1(0). \label{eq:boundary_eqn2}
\end{align}
Substituting Eq.~\eqref{eq:leadan} and Eq.~\eqref{eq:CIan} into Eq.~\eqref{eq:boundary_eqn1} and Eq.~\eqref{eq:boundary_eqn2} we can rewrite the boundary equations as,
\begin{equation}
	\begin{pmatrix}
		1 & -\eta_c/\eta_x\\
		\eta_ce^{iq} & c
	\end{pmatrix}
	\begin{pmatrix}
		r\\
		u
	\end{pmatrix}
	=
	-
	\begin{pmatrix}
		1\\
		\eta_ce^{-iq}
	\end{pmatrix},
\end{equation}
where 
\begin{equation}
	c=\frac{(\sin k-\epsilon)\lambda_2}{\lambda_2 + \mu_w + \cos k}.
\end{equation}

Solving this matrix equation gives,
\begin{align} \label{eq:reflection coefficient}
	r&=-\frac{
		c+\frac{\eta_c^2}{\eta_x}e^{-iq}
	}{
		c+\frac{\eta_c^2}{\eta_x}e^{iq}
	},\\
	\label{eq:u expression}
	u&=\frac{
		2i\eta_c\sin q
	}{
		c+\frac{\eta_c^2}{\eta_x}e^{iq}
	}.
\end{align}
Note that $e^{-iq}$ is related to the self-energy in the NEGF approach (see Sec.~\ref{section: NEGF} for more details) as
\begin{equation}
	\Sigma_{k}(\epsilon)=\frac{\eta_c^2}{\eta_x}e^{-iq} = \Lambda + i\gamma.
\end{equation}
where $\Lambda = \eta_c^2/\eta_x \cos q$ and $\gamma = -\eta_c^2/\eta_x \sin q$. So,
\begin{equation}
	r = -\frac{
		c+ \Lambda + i\gamma}{
		c+ \Lambda - i\gamma} = \frac{
		\gamma - i(c+ \Lambda)}{
		\gamma + i(c+ \Lambda)},
\end{equation}
Let, $z = \gamma + i(c+ \Lambda) = e^{\frac{i\phi}{2}}$, then
\begin{equation}
	r = \frac{z^{*}}{z} = e^{-i\phi},
\end{equation}
with
\begin{equation}
	\phi = 2\tan^{-1}\left(\frac{c+ \Lambda}{\gamma}\right)= -2\arctan\left[\frac{\eta_xc +\eta_c^2\cos q}{\eta_c^2\sin q}\right].
\end{equation}
As we will see in Sec.~\ref{section: winding number, reflection coefficient relation}, the winding number corresponding to $\phi(k)$ (as $k$ is varied in the allowed $k$-interval $\mathcal{I} \subseteq (-\pi, \pi)$) is equal to the Hall conductance of the lead-CI setup.
\subsection{Computing Hall conductance} \label{subsection: Hall conductance}

Having obtained the exact scattering state solution, we can move forward to compute our quantity of interest, Hall conductance. To compute it, we start by calculating the local current density along the $y$ direction using the following expression,
\begin{equation}
	j_y(x;q,k)=-2\text{Im}[\zeta_{q,k}^{\dagger}(x,y)H[x,y;x,y+1]\zeta_{q,k}(x,y+1)],
\end{equation}
where $H[x,y;x,y+1]$ is the hopping strength in the $y$- direction  in the CI or lead Hamiltonian.
Using the form of the wavefunction $\zeta_{q,k}(x,y)=\Psi_{q,k}(x)e^{i k y}$ and the matrix elements inside lead and CI, the $y$-current due to single particle state can be expressed as
\begin{equation} \label{eq: j_y}
	j_y(x ; q,k) = \Psi^{\dagger}_{q, k}(x)
	~\hat{J_y}~\Psi_{q, k}(x),
\end{equation}
where 
\begin{equation} \label{eq: current operators_scattering}
	\hat{J_y} = 
	\begin{cases} 
		\sigma_z \cos k - \sigma_x \sin k & \text{for } x \geq 0 ,\\
		-2\eta_y \sin k & \text{for }  x < 0 .
	\end{cases}
\end{equation}
is the current operator corresponding to the $y$-current. So, the total current due to  scattering states filled up to the Fermi level $\mu$ is given by,
\begin{equation} \label{eq: vertical current}
	J_y(\mu, x) = \frac{1}{2\pi} \int_0^{2\pi} dk \int_0^{q_F} dq~j_y(x ; q,k)~\Theta ( \mu - \epsilon_{q, k}),
\end{equation}
where $\epsilon_{q,k}$ is given by Eq.~\eqref{eq:dispersion}. Differentiating  Eq.~\eqref{eq: vertical current} with respect to $\mu$, we can define \textit{local} Hall conductance as,
\begin{equation} \label{eq: local conductance2}
	\tilde{G}_{H}(\mu, x) = \frac{1}{2\pi}\int_0^{2\pi} dk \int_0^{q_F} dq~j_y(x ; q,k)~\delta ( \mu - \epsilon_{q, k}).
\end{equation}
The total Hall conductance is then given by
\begin{equation}
	G_H(\mu) = G_H^{\text{lead}}(\mu) + G_H^{\text{CI}}(\mu),
\end{equation}
where
\begin{equation}  \label{eq:GHleaddef}
	G_H^{\text{lead}}(\mu) = \sum_{x=-\infty}^{-1} \tilde{G}_H^{\text{lead}}(\mu, x),
\end{equation}
is the contribution due to the leads and
\begin{equation} \label{eq: GHCIdef}
	G_H^{\text{CI}}(\mu)
	= \sum_{x=0}^{\infty} \tilde{G}_H^{\text{CI}}(\mu, x).
\end{equation}
is the contribution from the CI. Here, local Hall conductance can be obtained by computing the local $y$- current $j_y(x ; q,k)$, carried by single particle eigenstates and evaluating the $q$- integral in Eq.~\eqref{eq: local conductance2} first. This gives us
\begin{align}
	\tilde G_H^{\text{lead}}(\mu,x) &= \frac{1}{\pi} \int_{k\in\mathcal{I}} dk \frac{\eta_y\sin k~\cos (2qx + \phi(k))}{\eta_x\sin q},\label{eq:GHlmux1}\\
	\tilde G_H^{\text{CI}}(\mu,x) &= \frac{1}{\pi}\int_{k\in\mathcal{I}}dk\lambda^{x}\gamma\frac{(1 - f^2)\cos k - 2f \sin k}{\gamma^2 + (c+\Lambda)^2}.\label{eq:GHcimux2}
\end{align}
The local contributions from the CI can be summed easily by substituting Eq.~\eqref{eq:GHcimux2} in Eq.~\eqref{eq: GHCIdef} to obtain,
\begin{equation}
	G_H^{\text{CI}}(\mu) = \frac{1}{\pi}\int_{k\in\mathcal{I}}dk\frac{\gamma(k)}
	{\gamma(k)^2 + (c(k) + \Lambda(k))^2}h(k).\label{eq:GH_CI}
\end{equation}
where $h(k) = ((1 - f^2)\cos k - 2af)/(1-\lambda_1^2)$. Summing local contributions inside the leads is more subtle and is discussed in Sec.~\ref{section: sumGHleads}.

Fig.~2 of the main text shows how contributions from lead and CI add up to 1 in the topological regime. The numerical plots clearly indicate that quantization is retained by including the contribution from the leads. An analytic proof of this fact is discussed in the main text, with technical details given in Sec.~\ref{section: winding number, reflection coefficient relation}. 
\section{Hall conductance contribution from the leads} \label{section: Hall conductance contribution from the leads}
In this section, we analyse the decay behaviour of local Hall conductance inside the leads for three different parameter regimes given in the main text namely,
\begin{equation}
	\begin{array}{ll}
		\text{(I)} & \eta_x>\eta_y,\ \text{and}\ \mu<2\Delta\eta,~~~
		k\in(-\pi,\pi),\\[2pt]
		\text{(II)} & \mu>2\Delta\eta,\qquad
		~~~~~~~~~~~~~~~k\in(-k_2,k_2),\\[2pt]
		\text{(III)} & \eta_x<\eta_y,\ \text{and}\ \mu<2\Delta\eta,~~~
		k\in(-k_2,-k_1)\cup(k_1,k_2).
	\end{array}
	\label{eq:cases}
\end{equation}
where $\Delta \eta \equiv |\eta_x - \eta_y|,~k_1=\arccos[(\mu+2\eta_x)/(2\eta_y)]$, and $k_2=\arccos[(\mu-2\eta_x)/(2\eta_y)].$ These three different cases arise due to different allowed regions for the value of $k$ (see Fig.~\ref{fig:Fermi_surface} below) through the dispersion Eq.~\eqref{eq:dispersion}.
\begin{figure}[h]
	\centering
	\includegraphics[width=0.9\linewidth]{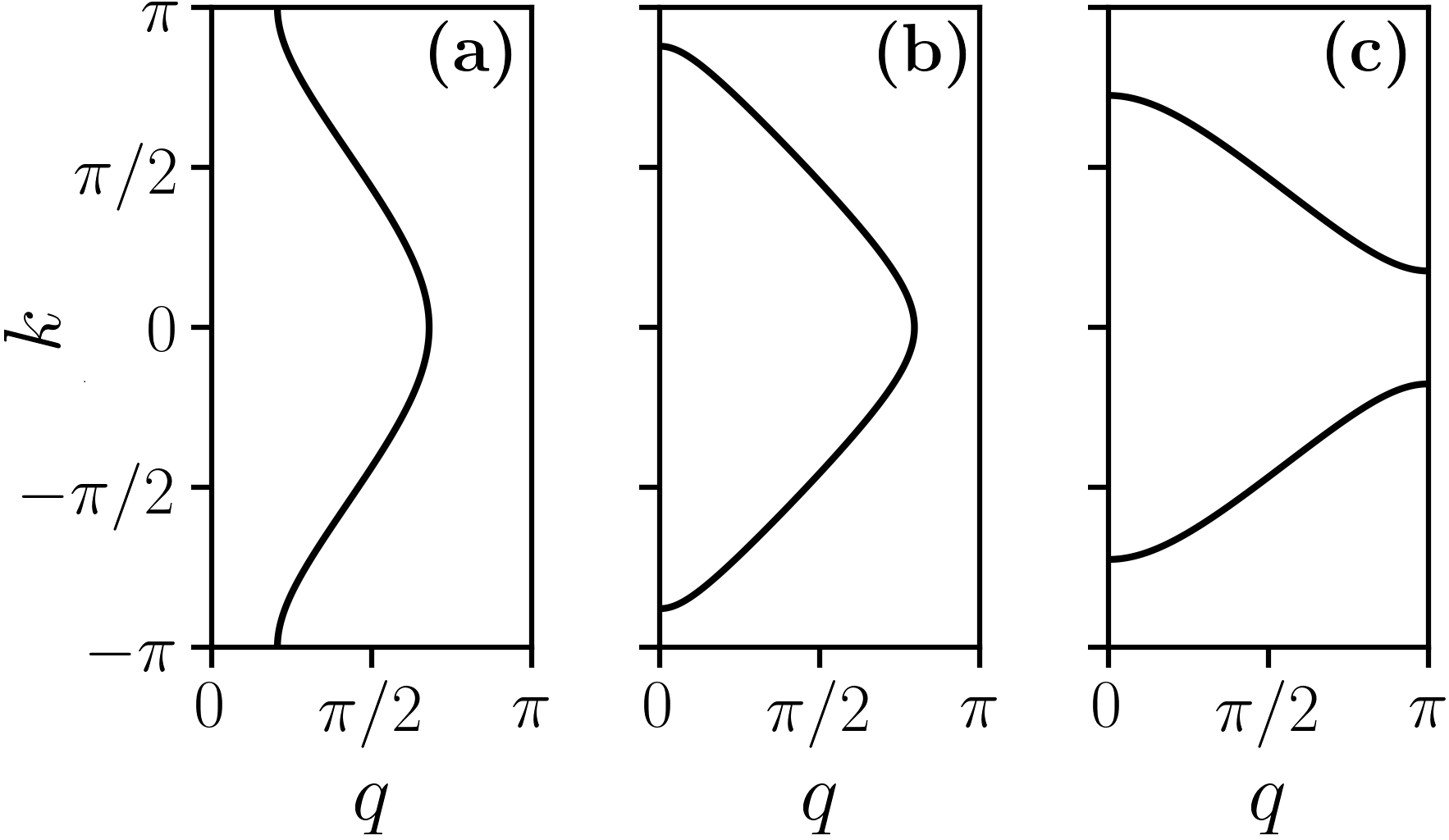}
	\caption{\textbf{Fermi surface of leads:~}The dispersion of leads results in three different Fermi surfaces for the three parameter regimes defined in Eq.~\eqref{eq:cases}. Panels (a)–(c) correspond to cases (I)–(III), respectively. The common parameters are $\eta_x=1.5$ and $\mu=0.5$ and other parameters, (a) $\eta_y=1$, $\Delta\eta=0.5$;
		(b) $\eta_y=1.4$, $\Delta\eta=0.1$;
		(c) $\eta_y=2$, $\Delta\eta=0.5$.}\label{fig:Fermi_surface}
\end{figure}

Later, we show that the contribution to the Hall conductance from this leaked Hall current can be summed nicely to obtain a closed-form expression.
\subsection{Decaying nature of local Hall conductance inside the leads}
To analyse the decaying nature of local Hall conductance, we use Eq.~\eqref{eq:GHlmux1}. As we will see, we obtain qualitatively different behaviours of decay of Hall conductance inside the leads for cases (I), (II), and case (III). 
\subsubsection{Case (I) and (II): Power law decay}
We first discuss case (I) and (II) mentioned in Eq.~\eqref{eq:cases}. To analytically obtain the decay nature for these two cases, we evaluate the local Hall conductance in the large $x$ limit using the stationary phase approximation. Consider Eq.~\eqref{eq:GHlmux1}, which can be rewritten as
\begin{equation}
	\tilde G_H^{\text{lead}}(\mu,x)
	=
	\Re\!\left[
	\int_{-\pi}^{\pi}
	dk\,A(k)e^{2ixq_F(k)}
	\right],~~
	A(k)=F(k)e^{i\phi(k)}.
\end{equation}
where $F(k) = (\eta_y\sin k)/(\pi \eta_x \sin q)$.
The stationary points are determined by finding values of $k$ where the derivative of the phase of $e^{2ixq_F(k)}$ vanishes. So,
\begin{equation}
	\frac{dq}{dk}
	= -\frac{\eta_y\sin k}{\eta_x\sin q} =0,
\end{equation}
gives $k=0, \pi$. For simplicity, we will focus on the case (II) for which the allowed range of $k$ is restricted to $(-k_2, k_2)$ where $k_2 = \cos^{-1}(\frac{\mu-2\eta_x}{2\eta_y})$. As a result, only the $k=0$ stationary point can contribute. The following calculation also holds for case (I) with the only difference that contributions from both the stationary points $k=0$ and $k=\pi$ need to be taken.

Expanding $q_F(k)$ present in the exponential phase of the integrand, around $k=0$, we obtain
\begin{equation}
	q_F(k)
	=
	q_0
	+\frac12q_0''k^2
	+
	O(k^4),
\end{equation}
where $q_0=q(0)$ and $q_0''=q''(0)$. Similarly, expanding the amplitude,
\begin{equation}
	A(k)
	=
	A(0)
	+
	A'(0)k
	+
	\frac12A''(0)k^2
	+
	O(k^3).
\end{equation}

Since $A(0) = F(0)e^{i\phi(0)}$ and $F(0) = -q'(0)/ \pi$,
we have $A(0) = 0$. As a result,
\begin{equation}
	\tilde G_H^{\text{lead}}(\mu,x) \approx
	\Re\!\int_{-\delta}^{\delta}\!dk\,
	\bigl(A'(0)k+\tfrac12A''(0)k^2+\cdots\bigr)
	e^{ix(2q_0+q_0''k^2)},
\end{equation}
where $\delta$ is a very small non-zero number. The stationary-phase approximation then implies that the contribution from a neighbourhood of the stationary point in the limit $|x| \rightarrow \infty$ is asymptotically equivalent to integral performed over the full real line,
\begin{equation}
	\tilde G_H^{\text{lead}}(\mu,x)\approx
	\Re\!\int_{-\infty}^{\infty}\!dk\,
	\bigl(A'(0)k+\tfrac12A''(0)k^2+\cdots\bigr)
	e^{ix(2q_0+q_0''k^2)}.
\end{equation}
Here, the first term vanishes identically because the integrand is an odd function. Hence, the leading contribution to local Hall conductance is given by,
\begin{equation}
	\tilde G_H^{\text{lead}}(\mu,x)
	\approx
	\Re\!\left[
	\frac{A''(0)}{2}
	e^{2ixq_0}
	\int_{-\infty}^{\infty}
	k^2
	e^{ixq_0''k^2}
	\,dk
	\right].
\end{equation}

\begin{figure}
	\centering
	\includegraphics[width=0.9\linewidth]{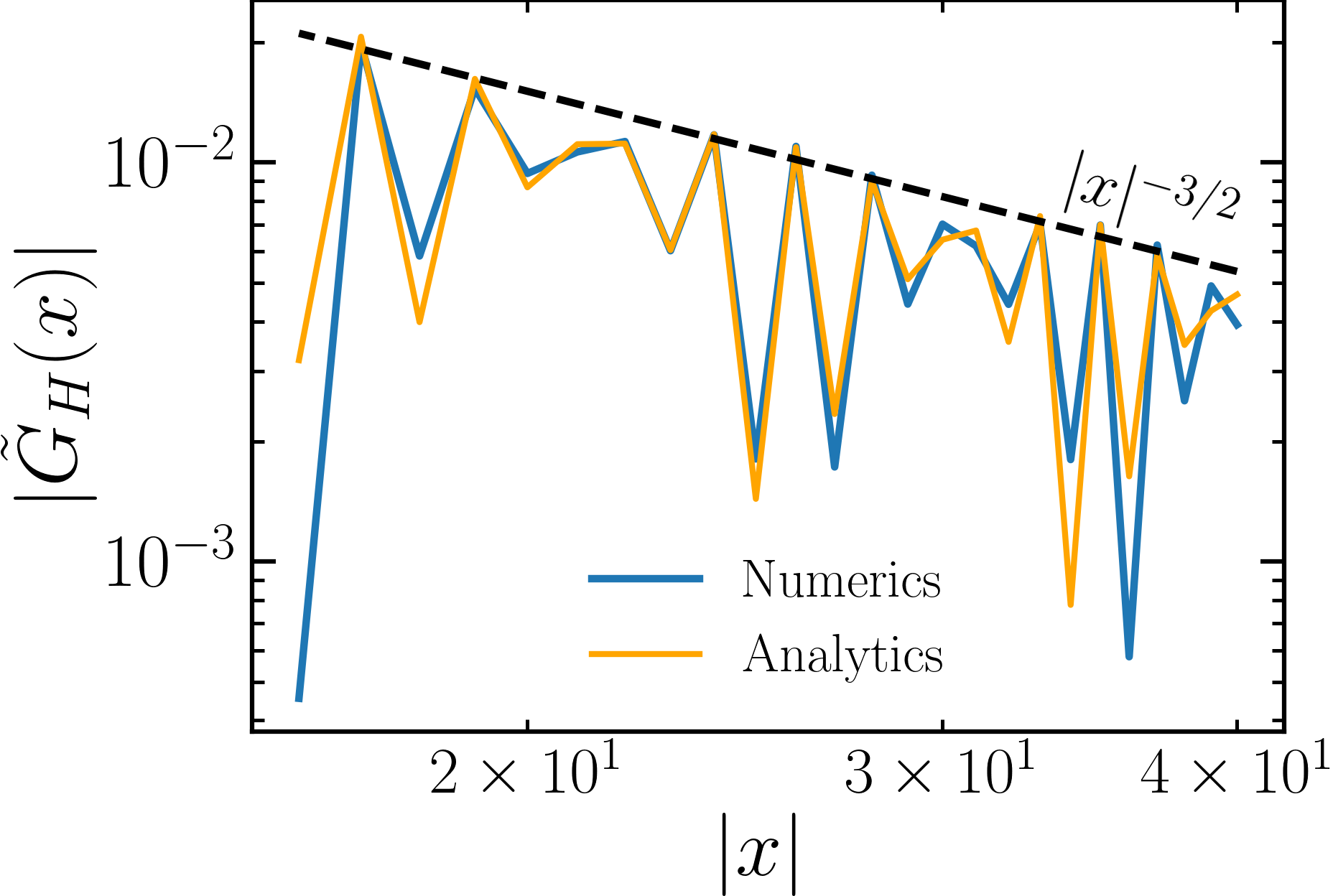}
	\caption{\textbf{Power-law decay}: Log-log plot of the local Hall conductance as a function of $x$ for case (II) defined by the condition $\mu>2\Delta\eta$. Parameter values: $\mu_w=1.1, \mu=0.5,~\eta_x=1.5,~\eta_y = 1.4,~\eta_c = 3$.}
	\label{fig:powerlaw}
\end{figure}

Using the standard oscillatory Gaussian integral, we obtain,
\begin{equation}
	\tilde G_H^{\text{lead}}(\mu,x)
	\approx
	\frac{1}{|x|^{3/2}}\frac{A''(0)\sqrt{\pi}\,
		\cos (2q_0x)}{
		4\,|q''(0)|^{3/2}\
	} \sim
	\frac{\cos(2q_0x)}{|x|^{3/2}}
\end{equation}
The power law decay and the oscillation frequency obtained above from the stationary phase approximation are compared and checked with numerics using Eq.~\eqref{eq:GHlmux1}. The comparison is shown in Fig.~\ref{fig:powerlaw} and we find a very good agreement between analytics and numerics. Small deviations in the plot are due to inaccuracies arising from a highly oscillating integrand in Eq.~\eqref{eq:GHlmux1} and since the approximation holds only for larger $x$.
\subsubsection{Case (III): Exponential decay}

Here, we treat case (III) separately, as the allowed range of $k$, $(-k_2,-k_1) \cup (k_1,k_2)$, does not include the stationary points $k=0, \pi$. As a result, here we express local conductance as a $q$-integral and convert it to a contour integral to obtain its decay nature, which turns out to be exponential in nature.

We start from the basic equation for the local conductance, Eq.~\eqref{eq: local conductance2}. Using the ansatz $\Psi_{q,k}(x) = e^{iqx} + r\,e^{-iqx}$ and the current operator $\hat{J_y} = -2 \eta_y\sin k$  in the lead region $(x < 0)$, we obtain,
\begin{equation}\label{eq: local conductance explicitt}
	\tilde{G}_{H}^{\text{lead}}(\mu,x)
	=\frac{2\eta_y}{\pi}
	\int_0^{\pi}\!\!dq
	\int_{-\pi}^{\pi}\!\!dk\,
	\sin k\,
	\cos(2qx+\phi)\,
	\delta(\mu-\epsilon_{q,k}).
\end{equation}
Note that for case (III), the Fermi surface splits into two branches (see Fig.\ref{fig:Fermi_surface}c). The equation for the two branches can be obtained using the Fermi surface equation $2\eta_x \cos q + 2\eta_y \cos k = \mu$ which gives,
\begin{equation} \label{eq: branches}
	k_{\pm}(q) = \pm \cos^{-1}\left(\frac{\mu - 2\eta_x \cos q}{2\eta_y}\right).
\end{equation}
\begin{figure}[h]
	\centering
	\includegraphics[width=0.88\linewidth]{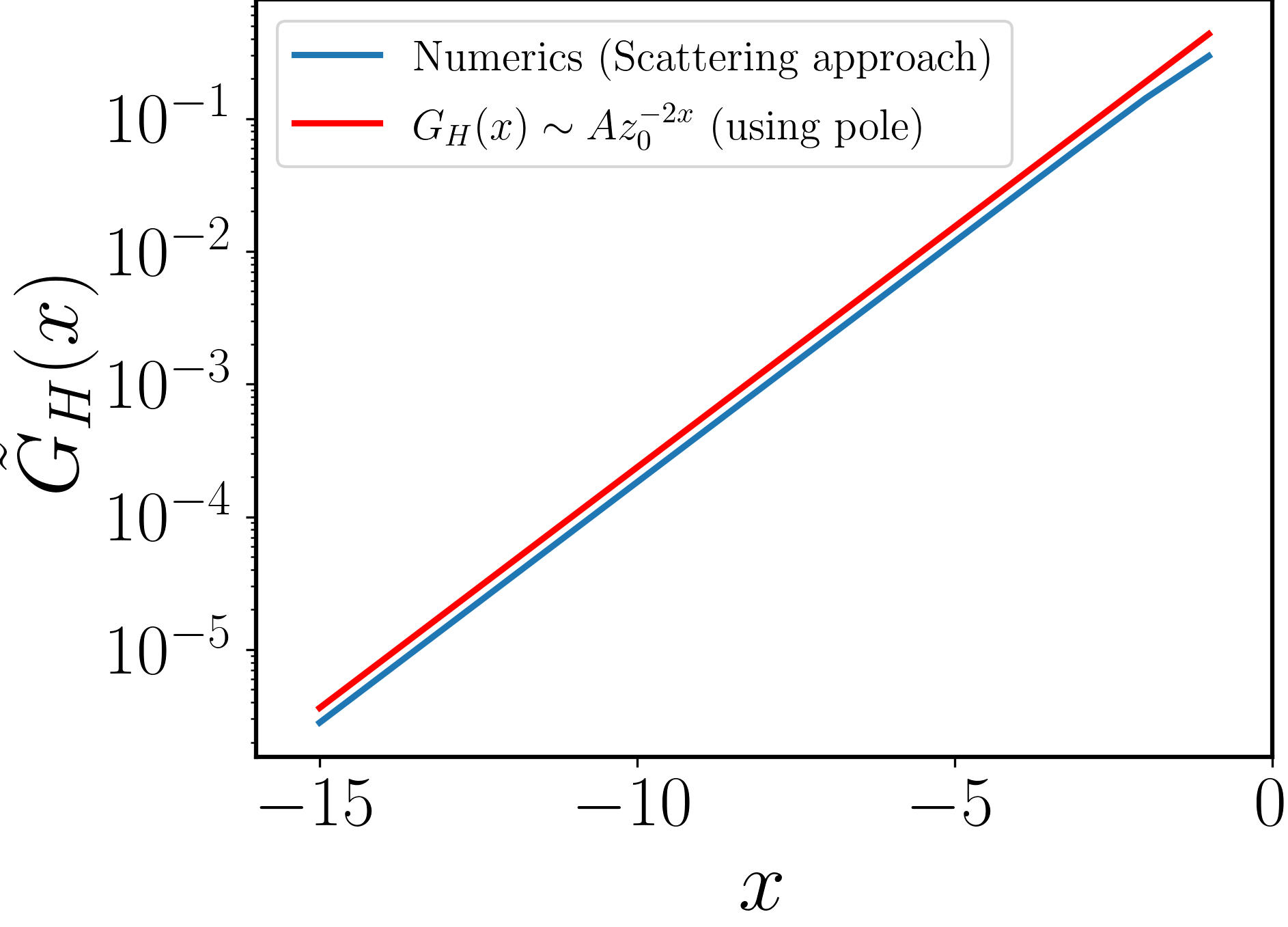}
	\caption{\textbf{Exponential decay: }Log-linear plot of Local Hall conductance as a function of $x$. The pole at $ z=0.6587$ gives the leading exponential behaviour. Parameter values: $\mu_w=1.001, \mu=0,~\eta_x=1,~\eta_y = 2,~\eta_c = 1.2$.}
	\label{fig:exponentialdecay}
\end{figure}
Now, using these expressions for two branches, we can evaluate the $k$ integral in Eq.~\eqref{eq: local conductance explicitt} to get,
\begin{align}
	\tilde G_H^{\text{lead}}(\mu,x)
	&= \frac{2\eta_y}{\pi}\int_0^{\pi} dq\,
	\Bigg[
	\frac{\sin k_{-}(q)}{|\sin k_{-}(q)|}
	\cos\!\left(2qx+\phi(q,k_{-})\right)
	\nonumber\\
	&\qquad\qquad
	+\frac{\sin k_{+}(q)}{|\sin k_{+}(q)|}
	\cos\!\left(2qx+\phi(q,k_{+})\right)
	\Bigg],\\
	&= \frac{2\eta_y}{\pi}\int_{0}^{\pi} dq\,
	\Big[
	\cos\!\left(2qx+\phi(q,k_{+})\right)
	\nonumber\\
	&\qquad\qquad
	-\cos\!\left(2qx+\phi(q,k_{-})\right)
	\Big].
\end{align}
which can be further simplified using the property $\phi(-q,k) = -\phi(q,k)$, as follows:
\begin{equation} 
	\tilde G_H^{\text{lead}}(\mu,x) =  \frac{\eta_y}{\pi}\int_{-\pi}^{\pi} dq\left[ r(q, k_{+}) e^{-i2qx} -r(q, k_{-})e^{-i2qx}\right].
\end{equation}
where,
\begin{equation}
	r(q, k_{\pm}(q)) = \frac{-(c(k_{\pm})e^{iq} + \frac{\eta_c^2}{\eta_x})}{c(k_{\pm})e^{-iq} + \frac{\eta_c^2}{\eta_x}}.
\end{equation}
Converting these expressions into closed contour integrals through the substitution $z= e^{iq}$, we get
\begin{equation}
	\tilde G_H^{\text{lead}}(\mu,x)  = \frac{-i \eta_y}{\pi} \left[ \oint_{\Gamma} dz~f_1(z)~z^{-2x}  + \oint_{\Gamma} dz~f_2(z)~z^{-2x} \right],
\end{equation}
where the contour $\Gamma$ is taken counterclockwise along the unit circle $|z|=1$ in the complex $z$-plane with
\begin{equation}
	f_i(z) = \frac{zc_{i}(z) + \Bar{\gamma}}{c_{i}(z) + z\Bar{\gamma}}.
\end{equation}
Since it is difficult to analytically find the poles of the function $f_i(z)$ due to its complicated structure, we can numerically obtain the solutions of the equation:
\begin{equation}
	c_{i}(z) + z\Bar{\gamma} = 0.
\end{equation}

This equation gives the solutions to be $z = 0.3047$ and $0.6587$ for the parameters given in Fig.~\ref{fig:exponentialdecay}. The dominating zero at $z = 0.6587$
gives the leading exponential behaviour and has very good agreement with the decay length obtained using the scattering approach (See Fig.~\ref{fig:exponentialdecay}).
\subsection{Summing local Hall conductance inside the leads} \label{section: sumGHleads}
The Hall conductance contribution from the leads can be calculated as described in Sec.~\ref{subsection: Hall conductance}. To sum up all the contributions, we can rewrite the total lead contribution using Eq.~\eqref{eq:GHleaddef} and Eq.~\eqref{eq:GHlmux1} as: 
\begin{equation} \label{eq: GHleadlocal_compact}
	G_H^{\text{lead}}(\mu) =  \int_{k\in\mathcal{I}} dk ~F(k)\times \sum_{x=-\infty}^{-1} \cos (2qx + \phi(k)), 
\end{equation}
where $F(k) = \eta_y\sin k/ (\pi \eta_x\sin q)$.

We can split the integral in two terms as follows:
\begin{equation} \label{eq: GHlead=I1+I2}
	G_H^{\text{lead}}(\mu)
	= I_1 - I_2,
\end{equation}
where
\begin{align}
	I_1 &= \int_{k\in\mathcal{I}} dk \, F(k) \cos \phi(k)
	\sum_{x=-\infty}^{-1} \cos (2qx),\label{eq: I1}\\
	I_2 &= \int_{k\in\mathcal{I}} dk \, F(k) \sin \phi(k)
	\sum_{x=-\infty}^{-1} \sin (2qx).
\end{align}
The infinite sum present in $I_1$ can be alternatively expressed using Poisson summation as,
\begin{equation} \label{eq: Poisson summation}
	\sum_{x=-\infty}^{-1} \cos (2qx) = \frac{\pi}{2}\sum_{n=-\infty}^{\infty} \delta(q-\pi n) - \frac{1}{2}.
\end{equation}
Since $q \in [0, \pi]$, $\sum_{n=-\infty}^{\infty} \delta(q-\pi n) = \delta(q) + \delta(q- \pi)$. Physically, $q=0$ and $q=\pi$ points are not expected to contribute to the conductance, as for such points, $r=-1$ and so the wavefunction inside the leads vanishes. This can also be shown more precisely by substituting Eq.~\eqref{eq: Poisson summation} in Eq.~\eqref{eq: I1}, where depending on the allowed values of $q$ via the dispersion, such points are either not allowed or they come up in pairs and their contribution cancels. As a result, we have
\begin{equation}
	I_1 = -\frac{1}{2}\int_{k\in\mathcal{I}} dk \, F(k) \cos \phi(k). \label{eq: I1 final}
\end{equation}
Similarly, the infinite sum in $I_2$ expression can be simplified as follows,
\begin{align}
	\sum_{x=-\infty}^{-1} \sin (2qx)  &= \sum_{x=1}^{\infty} -\sin (2qx),\\
	&= \frac{1}{2}\frac{d}{d q} \sum_{x=1}^{\infty} \frac{\cos (2qx)}{x}, \label{eq: infinite sum}\\
	&= -\frac{1}{2}\frac{d}{d q}\text{ln}\left(2\sin q \right) = -\frac{1}{2}\cot q. \label{eq: summed}
\end{align}
Note that, to go from Eq.~\eqref{eq: infinite sum} to Eq.~\eqref{eq: summed}, we have used a known identity (see Ref. \cite{arfken}). As a result, we can write,
\begin{equation}
	I_2 = -\frac{1}{2}\int_{k\in\mathcal{I}} dk \, F(k) \sin \phi(k) \cot q \label{eq: I2 final}
\end{equation}
Using \eqref{eq: I1 final} and \eqref{eq: I2 final}, the expression for $G_H^{\text{lead}}(\mu)$ \eqref{eq: GHlead=I1+I2} can be compactly written as,
\begin{equation}
	G_H^{\text{lead}}(\mu) =-\frac{1}{2}\int_{k\in\mathcal{I}} F(k) \frac{\sin (q - \phi(k))}{\sin q} ~ dk.\label{eq: GH_lead_summed}
\end{equation}
\section{Winding number and its relation to reflection coefficient} \label{section: winding number, reflection coefficient relation}
In this section, we prove a crucial property of the function $h(k)$ and $c(k)$ present in the $G_H^{\text{CI}}$ expression and that the Hall conductance obtained by summing the contributions from both leads and CI gives us a winding number expression.\\[2mm]
\subsection{Relation between $c(k)$ and $h(k)$} \label{section:relation between c and h}
Here, we show that $dc(k)/dk =h(k)$, a crucial relation used in the main text to express Hall conductance as a winding number. To start, recall that
\begin{align}
	c(k)&=\frac{a-\mu}{1+b\lambda_1} \label{eq: c(k)},\\[1mm]
	h(k)&=((1 - f^2)\cos k - 2af)/(1-\lambda_1^2), \label{eq: h(k)}
\end{align}
where
\begin{equation}
	f(k)=\frac{\mu-a}{b+1/\lambda_1} = \frac{\lambda_1 + b}{\mu + a}. \label{eq: f(k)}
\end{equation}
Note that here all variables except $\mu$ depend on $k$, but for notational simplicity, we will suppress all $k$ dependence throughout the derivation. Using the above expressions, it can be shown that 
\begin{equation}
	f = -\lambda_1 c. 
\end{equation}
Using this equation, $h(k)$ can be expressed as \begin{equation}
	h(k)=
	\frac{
		\cos k\bigl(1-\lambda_1^2 c^2\bigr)
		+2a\lambda_1c
	}
	{1-\lambda_1^2}.
\end{equation}
Now, the derivative of $c(k)$ is given by,
\begin{equation}
	\frac{dc}{dk}
	=
	\frac{\cos k+a\lambda_1 c-bc\,\lambda_1'}
	{1+b\lambda_1},
	\label{eq:dc1}
\end{equation}
where, using the definitions from the glossary, it can be shown that,
\begin{equation}
	\lambda_1'
	=
	\frac{a\lambda_1^2}
	{b(\lambda_1^2-1)}
	\Bigl[\lambda_1 + 1/ \lambda_1 + 2(b-\cos k)\Bigr].
	\label{eq:lambda1prime}
\end{equation}
Substituting Eq.~\eqref{eq:lambda1prime} in Eq.~\eqref{eq:dc1} and simplifying, we get
\begin{equation}
	\frac{dc}{dk}
	=
	\frac{\cos k(\lambda_1^2 + 2ac\lambda_1^2 - 1) - 2a\lambda_1 c(1 + b\lambda_1)}
	{(1+b\lambda_1)(\lambda_1^2-1)}. \label{eq: dc/dk}
\end{equation}
To simplify the expression further, we will factor out $(1+b\lambda_1)$ from the numerator.\\[1mm]
For this, we note that using the definition $f(k)$ and $c(k)$ from Eq.~\eqref{eq: f(k)}, Eq.~\eqref{eq: c(k)} and the relation between them $f = -c\lambda$, we can write
\begin{align}
	\lambda_1 + b &= f(\mu + a)\\
	&= f(2a - c(1 +b\lambda_1))\\
	&=-2ac\lambda_1 + c^2\lambda_1(1 + b\lambda_1)
\end{align}
Multiplying $\lambda_1$ on both sides and rearranging, we get
\begin{equation}
	\lambda_1^2 + 2ac\lambda_1^2 = c^2\lambda_1^2(1+b\lambda_1) - b\lambda_1
\end{equation}
The above equation can be directly substituted in the first term of the numerator of Eq.~\eqref{eq: dc/dk}, which gives
\begin{equation}
	\frac{dc}{dk}
	=
	\frac{\cos k(c^2\lambda_1^2(1+b\lambda_1) - (b\lambda_1 + 1)) - 2a\lambda_1 c(1 + b\lambda_1)}
	{(1+b\lambda_1)(\lambda_1^2-1)}
\end{equation}
Now, we can factorize $-(1 + b\lambda_1)$ and this gives us the desired result,
\begin{equation}
	\frac{dc}{dk} = \frac{\cos k(1 - \lambda_1^2c^2) + 2a\lambda_1 c}{1 - \lambda_1^2} = h(k)
\end{equation}
\subsection{Expressing Hall conductance in terms of the winding of the reflection coefficient}
The Hall conductance contribution from CI is given by Eq.~\eqref{eq:GH_CI}, which can be rewritten as:
\begin{equation}
	G_H^{\text{CI}}(\mu)  = \frac{1}{\pi} \int_{k\in\mathcal{I}} dk\frac{1}{ 1 + (\frac{c(k) + \Lambda(k)}{\gamma(k)})^2} \frac{h(k)}{\gamma} \, dk, \label{eq:GH_CI1}
\end{equation}
where $h(k) = ((1 - f^2)\cos k - 2af)/(1-\lambda_1^2)$.\\

In Eq.~\eqref{eq:GH_CI1},  we substitute $t = \frac{c(k) + \Lambda(k)}{\gamma(k)}$ and using the result $\frac{dc(k)}{dk} =h(k)$ from Sec.~\ref{section:relation between c and h}, we obtain
\begin{equation}
	G_H^{\text{CI}}(\mu) =\int_{k\in\mathcal{I}}\frac{dk}{2\pi} \frac{d\phi(q_F, k)}{dk}-\int_{k\in\mathcal{I}}\frac{dk}{\pi} \frac{\Lambda' \gamma - \gamma' (c + \Lambda)}{\gamma^2 + (c + \Lambda)^2}.
\end{equation}
Using the definition of $\Lambda$ and $\gamma$ and their property that $\Lambda'=\gamma q'$ and $\gamma'=-\Lambda q'$, the expression can be written as,
\begin{equation} \label{eq:G_H_CI_sumbefore}
	G_H^{\text{CI}}(\mu) =\int_{k\in\mathcal{I}}\frac{dk}{2\pi} \frac{d\phi(q_F, k)}{dk}-\int_{k\in\mathcal{I}}\frac{dk}{2\pi}~q' \frac{2\gamma^2+2\Lambda(c+\Lambda)}{\gamma^2 + (c + \Lambda)^2}.
\end{equation}
Similarly, the Hall conductance contribution from the lead,
\begin{equation}
	G_H^{\rm lead}
	=
	-\frac{1}{2}
	\int_{k\in \mathcal{I}}
	dk~
	F(k)
	\frac{\sin(q-\phi)}{\sin q},\label{eq:Gls}
\end{equation}
can be simplified using the identity $\sin(q-\phi) = \sin q \cos \phi - \cos q \sin \phi$ and then expressing $\cos \phi$ and $\sin \phi$ in terms of $\tan \frac{\phi}{2}$. Further noticing that
\begin{equation}
	F(k) = \frac{1}{\pi}\frac{\eta_y\sin k}{\eta_x\sin q} = -\frac{1}{\pi}\frac{dq}{dk},
\end{equation}
and using the relation $\tan \frac{\phi}{2} = \frac{c + \Lambda}{\gamma}$, the expression can be rewritten as,
\begin{equation}\label{eq:G_H_lead_sumbefore}
	G_H^{\rm lead}
	=
	\frac{1}{2\pi}
	\int_I
	dk\,
	q'
	\frac{
		\gamma^2+\Lambda^2-c^2
	}
	{\gamma^2+(c+\Lambda)^2}.
\end{equation}
Adding Eq.~\eqref{eq:G_H_CI_sumbefore} and Eq.~\eqref{eq:G_H_lead_sumbefore}, we get
\begin{align}
	G_H^{\text{total}} &= G_H^{\text{CI}} + G_H^{\text{lead}}\\
	&= \frac{1}{2\pi}\int_{k\in\mathcal{I}} \frac{d\phi(k)}{dk}dk + \frac{1}{2\pi}\int_{k\in\mathcal{I}}dk~q' \frac{-\gamma^2 - (c + \Lambda)^2}{\gamma^2 + (c + \Lambda)^2},\\[4mm]
	&= \frac{1}{2\pi}\int_{k\in\mathcal{I}} \frac{d\phi(k)}{dk}dk - \frac{1}{2\pi}\int_{k\in\mathcal{I}}~\frac{dq}{dk}dk.
\end{align}
Since $q_F(k)$ has the same value at the endpoints of the interval $\mathcal{I}$~(for case (I)) or cancels each other's contribution (for case (II) and (III)), the second term never contributes. Therefore, we get,
\begin{equation}
	G_H(\mu) = G_H^{\text{total}}(\mu) = \frac{1}{2\pi}\int_{k\in\mathcal{I}} \frac{d\phi(k)}{dk}dk
\end{equation}
\section{Non-equilibrium Green's function (NEGF) formalism}
\label{section: NEGF}
In this section, we discuss the non-equilibrium Green's function (NEGF) formalism for setups in cylindrical geometry and outline the method used in Fig. 2 of the main text to numerically obtain the local Hall conductance inside the lead and CI.

In the NEGF approach~\cite{ref_dhar2006}, the non-equilibrium steady state (NESS) properties of the lead-CI setup can be expressed in terms of the effective single-particle Green's function of the system \cite{ref_dhar2006},
\begin{equation} \label{eq: effective Green's function}
	\mathcal{G}(\omega) = \frac{1}{\omega - H^S - \Sigma^L(\omega) - \Sigma^R(\omega)},
\end{equation}
where $H^S$ is the single-particle Hamiltonian of the isolated system, and $\Sigma^{L(R)}(\omega)$ represents the self-energy due to the left (right) lead. The Green's function is block diagonal in the basis given in Eq.~(\ref{eq:Fourier_transform_in_ydirection}) and therefore its elements can be written as,
\begin{align}
	\mathcal{G}_{x,a,m;x',b,m'}(\omega)&=\tilde{\mathcal{G}}^{m}_{x,a;x',b}(\omega)\delta_{m,m'},
\end{align}
where,
\begin{align}
	\Tilde{\mathcal{G}}^{m}&=\frac{1}{\omega-\Tilde{H}^{S}_m-\tilde \Sigma_m^{L}(\omega)-\Tilde{\Sigma}_m^{R}(\omega)}\label{eq:1D_GF},  
\end{align} 
and  $\tilde \Sigma_m^{L}(\omega)$, $\tilde \Sigma_m^{R}(\omega)$ are the self energies of the decoupled chains: 
\begin{eqnarray}\label{eq:sigmal}
	[\tilde \Sigma_m^{L}]_{x,s; x',s'} &=& \Sigma_m(\omega)\delta_{x,1}\delta_{x',1}\delta_{s, 1}\delta_{s',1},\\\label{eq:sigmar} 
	[\tilde \Sigma_m^{R}]_{x,s; x',s'} &=& \Sigma_m(\omega)\delta_{x,N_x}\delta_{x',N_x}\delta_{s,2}\delta_{s',2}.
\end{eqnarray}
$\Sigma_m(\omega)$ is defined as follows,
\begin{align} \label{eq:selfenergy}
	\Sigma_m(\omega)=\begin{cases}
		\frac{{\eta_c}^2}{\eta_x} \left[ \frac{\bar{\omega}_m}{2\eta_x} + \sqrt{\frac{\bar{\omega}_m^2}{4\eta_x^2}- 1} \right], & \text{if}~\bar{\omega}_m < -2\eta_x,\\
		\frac{{\eta_c}^2}{\eta_x} \left[ \frac{\bar{\omega}_m}{2\eta_x} - i\sqrt{1 - \frac{\bar{\omega}_m^2}{4\eta_x^2}} \right], & \text{if}~|\bar{\omega}_m| < 2\eta_x,\\
		\frac{{\eta_c}^2}{\eta_x} \left[ \frac{\bar{\omega}_m}{2\eta_x} - \sqrt{\frac{\bar{\omega}_m^2}{4\eta_x^2}- 1} \right], & \text{if}~\bar{\omega}_m > 2\eta_x,
	\end{cases}
\end{align}
with $\bar{\omega}_m = \omega - 2\eta_y \cos k_m$.

The steady-state two-point correlation matrix of the system can be expressed as \begin{align}
	C_{x, y,s;x', y',s'}&=\langle c^{\dagger}_s(x,y)c_{s'}(x',y')\rangle,\label{eq:corr2}
\end{align}
where ⟨·⟩ denotes the expectation value in the non‐equilibrium steady state. Using Eq.~(\ref{eq:Fourier_transform_in_ydirection}), the correlation matrix can be written as,
\begin{eqnarray}\non
	C_{x, y,s;x', y',s'} &=&\frac{1}{N_y}\sum_{m,m'=-{N_y}/{2}}^{N_y/2 -1}e^{i(k_my-k_{m'}y')}\langle\Tilde{c}^{\dagger}_{s,m}(x)\Tilde{c}_{s',m'}(x')\rangle,\\
	&=& \frac{1}{N_y}\sum_{m=-{N_y}/{2}}^{N_y/2 -1} e^{ik_m(y -y')}\Tilde{C}_{x,s;x',s'}^{m},\label{eq:Fourier_transformed_corrmat}
\end{eqnarray}
where we have used the fact that the Fourier modes along the $y$-direction are uncorrelated and hence $\langle\Tilde{c}^{\dagger}_{a,m}(x)\Tilde{c}_{b,m'}(x')\rangle= \Tilde{C}_{x,a;x',b}^{m}\delta_{m,m'}$. We note that $\tilde{C}^m$ is the correlation matrix for the effective 1D chains and can be expressed as~\cite{ref_dhar2006},
\begin{equation} \label{eq:Ctilde_m}
	\tilde C^m =\int_{-\infty}^{\infty}d\omega[\mathcal{A}^{ m,L}(\omega)f^{L}(\omega)+\mathcal{A}^{ m,R}(\omega)f^{R}(\omega)],
\end{equation} 
where
\begin{equation}
	\mathcal{A}^{m,L/R}(\omega) = 2\pi \tilde{\mathcal{G}}^{m}(\omega)\tilde{\Gamma}^{L/R}_m(\omega)(\tilde{\mathcal{G}}^m(\omega))^{\dagger},\label{eq:Clambdam1}
\end{equation}
$\tilde\Gamma^{L/R}_m(\omega)=(\tilde\Sigma^{L/R}_m(\omega)-[\tilde\Sigma^{L/R}_m(\omega)]^\dagger)/(2\pi i)$ and 
$f^{L/R}(\omega)=f(\omega,\mu_{L/R}, T_{L/R})$ are the Fermi distribution functions of the two reservoirs.
The expression for the steady-state current along the bond connecting the sites $\mathbf{x}=(x,y)$ and $\mathbf{x'}=(x',y')$ follows  from the continuity equation and can be written as,
\begin{equation}
	J_{\mathbf{x},\mathbf{x'}} = -2
	\sum_{s,s'=1}^2\text{Im}\left\{ H^S[\textbf{x}, s; \mathbf{x'}, s'] C_{\mathbf{x},s;\mathbf{x}',s'}\right\}.\label{eq:curr1}
\end{equation}
We note here that while Eq.~\eqref{eq:curr1} can be used to compute the current in the CI, the same equation with a slight modification can also be used to compute the current inside the metallic region~\cite{Bhat2025}. The main point to note is that in the nonequilibrium steady state of the lead-CI-lead setup, the state of the leads has also evolved and therefore contains the full information about the scattering states of the full system. Therefore, we consider a lead-metal-CI-metal-lead setup where the Hamiltonian of the metals is now identical to that of the leads, and the coupling of the metal with the CI is identical to that of coupling of the leads with CI in our typical lead-CI-lead setup. In other words, we include a part of the leads and their contacts into the wire  Hamiltonian, $H^S$, and then use the NEGF formalism. The current densities in the two setups are identical because of the uniqueness of the NESS. We have used this idea in Fig.~(2) of the main text to compute the current density inside the leads and shown that it is identical to what we obtain from the scattering states of the lead-CI-lead system.

The edge modes are localized either at the left or at the right boundary of the cylinder. Hence, they do not carry any current along the $x$-direction, so we focus exclusively on the total current in the $y$-direction, which is given by
\begin{eqnarray}
	{J_y} &=& \sum_{x=0}^{N_x -1} J_{x, y; x, y+1},\\
	&=&
	-2\sum_{x=0}^{N_x -1} \sum_{s,s'=1}^2\text{Im}\left\{ H^S_{x, y,s; x, y+1,s'} C_{x, y,s; x, y+1,s'}\right\}. ~~~~~~\label{eq:Jy_}
\end{eqnarray}
The subscript on $J_y$ only labels the direction of the current, and it does not depend on $y$ because of the cylindrical symmetry of the system. 

Substituting Eq.~\eqref{eq:Ctilde_m} in Eq.~\eqref{eq:Fourier_transformed_corrmat} and using the result in Eq.~\eqref{eq:Jy_}, $J_y$ can be rewritten as~\cite{Bhat2025},
\begin{equation} \label{eq:Jy}
	J_y = \sum_{x=0}^{N_x-1}\sum_{m=-{N_y}/{2}}^{N_y/2 -1} \int_{-\infty}^\infty d\omega [F^{m,L}_{x,x}(\omega)f^L(\omega)+F^{m,R}_{x,x}(\omega)f^R(\omega)],
\end{equation}
where,
\begin{equation} \label{eq: F_lambda}
	F^{m, L/R}_{x,x}(\omega)=-\frac{1}{N_y}\sum_{s,s'=1}^2\text{Im}\left[e^{-ik_m}(\sigma_x + i\sigma_z)_{s,s'} \mathcal{A}^{m, L/R}_{x, s;  x, s'}\right].
\end{equation}

Throughout this paper, we work at zero temperature and in the linear response regime, i.e., $\mu_R = \mu$ and $\mu_L = \mu + \Delta \mu$. Then, the steady-state current expression in Eq.~\eqref{eq:Jy}  naturally decomposes into two parts, an equilibrium part $J_{eq}$ and an excess part $\Delta J_y$~\cite{Bhat2025},
\begin{equation}
	J_{y} = J_{eq} + \Delta J_y,
\end{equation}
where
\begin{equation} \label{eq:Equilibrium current}
	J_{eq} = \sum_{x=0}^{N_x-1}\sum_{m=-{N_y}/{2}}^{N_y/2 -1} \int_{-\infty}^{\mu} d\omega~[ F^{m,L}_{x,x}(\omega)+F^{m,R}_{x,x}(\omega)],
\end{equation}
and
\begin{equation} \label{eq:Excess_current}
	\Delta J_y =\sum_{x=0}^{N_x-1}\sum_{m=-{N_y}/{2}}^{N_y/2 -1} F^{m,L}_  
	{x,x}(\mu)\Delta\mu.
\end{equation}

In Eq.~\eqref{eq:Equilibrium current}, the integration from $-\infty$ to $\mu$ includes contributions from both bulk and edge states. The bulk states appear in left and right moving pairs whose currents cancel each other. Within the band gap, the two chiral edge modes at a fixed Fermi level $\mu$ (see Fig.~\ref{fig:CIspectrum}) are exponentially localized at the cylinder’s opposite boundaries and propagate with opposite chirality. Thus, their contributions to the equilibrium current also cancel each other, giving $J_{eq}=0$. Hence, only the non-equilibrium excess current $\Delta J_y$ driven by $\Delta \mu$ remains, which is given by Eq.~\eqref{eq:Excess_current}. This current can be used to define Hall conductance as,
\begin{equation} \label{eq:hcdef}
	G_H(\mu) =  \lim_{\Delta \mu \rightarrow 0}\frac{\Delta J_y}{\Delta \mu}.
\end{equation}

We use Eq.~(\ref{eq: F_lambda}) and Eq.~(\ref{eq:Excess_current}) in  Eq.~(\ref{eq:hcdef}) and evaluate the imaginary part to express the Hall conductance as,
\begin{equation} \label{eq:GH_1D}
	G_H(\mu) =\frac{1}{N_y}\sum_{x=0}^{N_x-1}\sum_{m=-{N_y}/{2}}^{N_y/2 -1}\sum_{s,s'=1}^2~ R_{s,s'}(k_m)~\mathcal{A}^{m, L}_{x, s;  x, s'}(\mu),
\end{equation}
where 
\begin{equation} \label{eq:Rm}
	R_{s,s'}(k_m) = (\sigma_z \cos k_m - \sigma_x  \sin k_m)_{s,s'}.
\end{equation}
and
\begin{equation}
	\mathcal{A}^{m,L}_{x,s;x,s'}(\mu) = 2\pi [\tilde{\mathcal{G}}^{m}(\mu)\tilde{\Gamma}^{L}_m(\mu)(\tilde{\mathcal{G}}^m(\mu))^{\dagger}]_{x,s;x,s'}.
\end{equation}
Using Eq.~\eqref{eq:GH_1D}, Hall conductance can be rewritten in terms of local Hall conductance as,
\begin{equation} \label{eq:GH_1D_last}
	G_H(\mu) =\sum_{x=0}^{N_x-1}\tilde{G}_H(x),
\end{equation}
where, 
\begin{equation} \label{eq:local_cond_NEGF}
	\tilde{G}_H(x) = \frac{1}{N_y}\sum_{m=-{N_y}/{2}}^{N_y/2 -1}\sum_{s,s'=1}^2~ R_{s,s'}(k_m)~\tilde C^{m, L}_{x, s;  x, s'}(\mu),
\end{equation}
and
\begin{equation}
	\tilde C^{m,L}(\omega) = 2\pi~ \tilde{\mathcal{G}}^{m}(\omega)\tilde{\Gamma}^{L}(\omega)(\tilde{\mathcal{G}}^m(\omega))^{\dagger}.\label{eq:Clambdam2}
\end{equation}
Note that the quantity $\tilde{G}_H(x)$ can be interpreted as local conductance. This quantity can also be calculated inside the lead region by including the Hamiltonian of the coupling and a finite region of leads in the system Hamiltonian $\tilde{H}^S$ present in Eq.~\eqref{eq:1D_GF}~(For related discussions, see \cite{Bhat2025}).

Another thing that needs to be taken care of is that the $R(k_m)$ matrix in Eq.~\eqref{eq:local_cond_NEGF} corresponds to the current operator $\frac{\partial H(\mathbf{k})}{\partial k_y}$, where $H(\mathbf{k})$
is the momentum–space Hamiltonian of the region under consideration (either the lead or the Chern insulator). Here, $\mathbf{k} = (k_x, k_y)$ denotes the wavevector components along the $x$ and $y$ direction, respectively. For   different regions, it is given by
\begin{equation} \label{eq: current operators_NEGF}
	R(k_m) = 
	\begin{cases} 
		-2\eta_y \sin k & \text{for }  x \leq -1\\
		\sigma_z \cos k - \sigma_x \sin k & \text{for } 0 \leq x \leq N_x-1,\\
		-2\eta_y \sin k &\text{for}~ x \geq N_x.\\
	\end{cases}
\end{equation}
Here, $x \leq -1$ and $x \geq N_x$ correspond to the lead region, whereas $0 \leq x \leq N_x-1$ is the CI region.
\section{Anomalous Topological Phases: Robustness and its origin}
In this section, we discuss the case where we find that the quantized Hall conductance is not equal to the bulk Chern number (we call them 'anomalous topological phase (ATP)').
\subsection{Robustness under coupling Hamiltonian modifications}
Here, we discuss the effect of modifying the way of coupling (of CI to leads) on the presence of ATPs.
\begin{figure}[h]
	\centering
	\includegraphics[width=1\linewidth]{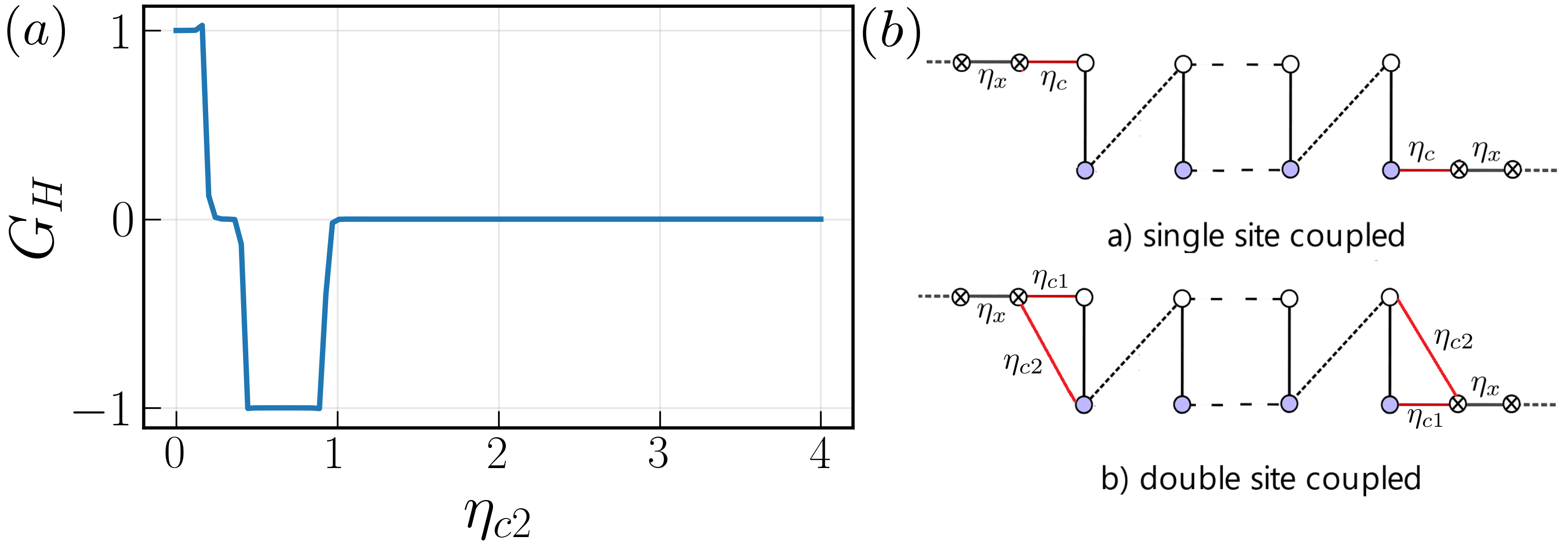}
	\caption{\textbf{Anomalous topological phase in a modified setup: }a) Hall conductance as a function of $\eta_{c2}$ in the non-topological regime for double site coupled setup. b) Schematic to show the different microscopic models for coupling to leads. Parameter values: $\eta_x = 1,\ \eta_y = 1.5,\ \eta_{c1} = 3,\ \mu_w = 2.5,\ N_x = 20,\ N_y = 10^4,\ N_x = 20$.}
	\label{fig:couplingchange}
\end{figure}
A natural question to ask is whether ATPs exist even when we couple the leads in a different manner to CI.
For example, we can modify the coupling Hamiltonian of the effective 1D model as follows:
\begin{align} 
	&\tilde{\mathcal{H}}_m^{SL} = \eta_{c1} \tilde c_{L,m}^\dagger(-1)\,\tilde c_{1,m}(0) + \eta_{c2} \tilde c_{L,m}^\dagger(-1)\,\tilde c_{2,m}(0) + \text{h.c.},
	\label{eq:couplingH1}\\
	&\tilde{\mathcal{H}}_m^{SR} = \eta_{c1} \tilde c_{2,m}^\dagger(N_x-1)\,\tilde c_{R,m}(N_x) \nonumber \\ 
	& ~~~~~~~~~~~~~~~+ \eta_{c2} \tilde c_{1,m}^\dagger(N_x-1)\,\tilde c_{R,m}(N_x) + \text{h.c.} .\label{eq:couplingH2}
\end{align}
These modifications in the Hamiltonian result in a double-site coupled setup where both degrees of freedom of CI are coupled to leads with different coupling strengths $\eta_{c1}$ and $\eta_{c2}$ as shown in Fig.~\ref{fig:couplingchange}b. As we can see from the figure, this model reduces to our old model in the limit $\eta_{c2} \rightarrow 0$ with $\eta_{c1}= \eta_c$.

To check whether we observe ATP in this modified setup, we plot the Hall conductance using direct NEGF numerics keeping $\eta _{c1}$ fixed and varying $\eta_{c2}$. As shown in Fig.~\ref{fig:couplingchange}a, we find that the Hall conductance can indeed take values $\pm 1$ and $0$ as we vary $\eta_{c2}$, leading to ATPs in the non-topological regime. This numerical plot demonstrates the robustness of our results to the microscopic modelling of coupling. 
\subsection{Origin of ATP in the weak coupling limit}
\begin{figure}[h]
	\centering
	\includegraphics[width=1\linewidth]{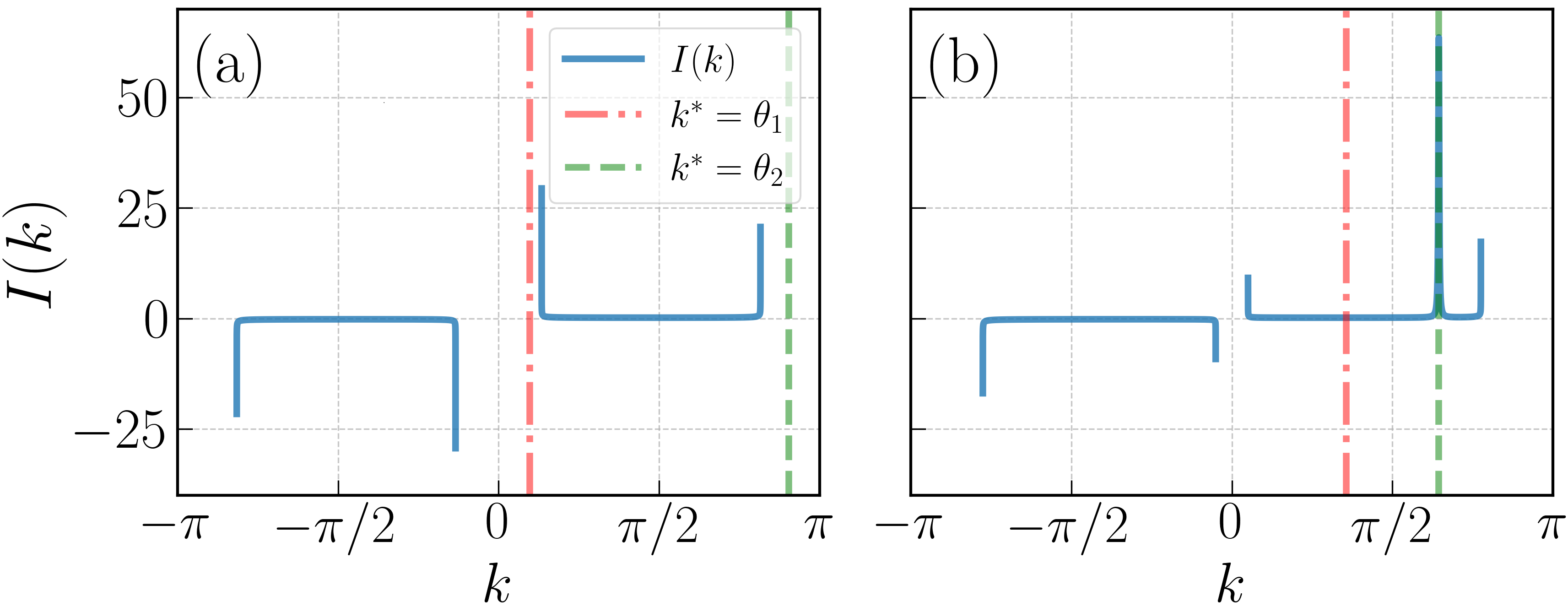}
	\caption{\textbf{Integrand of $G_H~$vs ~$k$ in the weak coupling limit.} (a) shows the integrand for $\mu=0.3$ and  (b) shows the integrand $\mu=0.9$. We see a distinctive peak in (b) as $\mu=\sin k$, the energy of the edge state, lies on the Fermi surface. This peak contributes to the integral in the weak coupling limit, and we get nonzero quantization of $G_H$. This peak is absent in (a) and therefore $G_H=0$ in the weak coupling limit. Other parameters: $\mu_w = 1, \eta_x = 3.5, \eta_y = 4, \eta_c = 0.12$, $\theta_1=\arcsin\mu$ and $\theta_2=\pi-\arcsin\mu$.}
	\label{fig:weakcoupling}
\end{figure}
In this section, we give a simple physical explanation for the presence of ATPs in our setup in the weak coupling limit.
It is typically expected that, in the limit of weak coupling~($\eta_c \rightarrow 0$) between lead and CI, we will recover the physics of the isolated CI. 
However, as evident in the phase diagram (Fig. 4) given in the main text, the Hall conductance vanishes in our setup even in the topological regime in the $\eta_c \rightarrow 0$ limit for a 
particular range of Fermi level, $\mu$. Note that these anomalous phases are present for certain regimes of the lead parameters ($\eta_x, \eta_y, \mu$) that correspond to case (II) and case (III) of Eq.~\eqref{eq:cases}. For these two cases, the allowed values of $k$ (electron's quasi-momentum along the $y$-direction) are constrained and do not span the full Brillouin zone.

 To understand the consequence of this, we focus on case (III), i.e., $\eta_x<\eta_y,\ \text{and}\ \mu<2\Delta\eta$, where the allowed values of $k$ inside the leads are restricted to $\mathcal{I} \in (-k_2,-k_1)\cup(k_1,k_2)$. Furthermore, translational invariance along the $y$-direction and fixed $\mu$ put kinematic constraints that both energy and $y$-momentum of the incoming electron must be conserved as the electron gets scattered at the Lead-CI junction. The dispersion for the edge states lying inside the band gap of the CI  is given by $\epsilon = \sin k$,  $k^{*} = \sin^{-1} \mu$, and $\pi - \sin^{-1} \mu$ should lie inside $\mathcal{I}$ for the edge modes to hybridize with the modes of the lead.  However, $k^{*}$ may not lie inside $\mathcal{I}$, and hence in such cases the incoming plane wave will not hybridize with the edge state of the CI. We infer this by looking at the integrand, $I(k)$, of $G_H$, which is the sum of integrands of Eq.~\eqref{eq:GH_CI} and Eq.~\eqref{eq: GH_lead_summed}. We show   $I(k)$ as a function of $k$ for $\mu=0.3$ and for $\mu=0.9$ in Fig. \ref{fig:weakcoupling}a and  Fig. \ref{fig:weakcoupling}b, respectively. For $\mu=0.9$, $k^{*}$ lies inside $\mathcal{I}$ leading to $G_H=1$ whereas for $\mu=0.3$, $k^{*}$ does not lie in $\mathcal{I}$ resulting in $G_H=0$. This is indicated by a distinctive peak at $k^*=\pi-\arcsin\mu$ in Fig.~\ref{fig:weakcoupling}b, which is absent in Fig.~\ref{fig:weakcoupling}a. This peak arises from the hybridization of the edge modes of the CI with the incoming plane waves from the lead.

\end{document}